\documentclass[a4paper]{article}
\usepackage[utf8]{inputenc}
\usepackage[textwidth=17.0cm,textheight=25cm]{geometry}
\usepackage{amsmath,amsfonts,amssymb,color,graphicx,epsf,subfigure}
\usepackage{authblk}
\usepackage{mathtools}
\usepackage[toc,page]{appendix}
\usepackage{verbatim}
\usepackage{cite}

\newcommand{\eq}[1]{(\ref{eq:#1})}  % referencing equations simplified
\newcommand{\fig}[1]{\textbf{\ref{fig:#1}}}  % referencing figures simplified
\newcommand{\Sec}[1]{\textbf{\ref{sec:#1}}}  % referencing sections simplified
\newcommand{\Op}{\mathcal{O}}  % operator O
\newcommand{\Tr}{\text{Tr}}  % trace
\newcommand{\ket}{\rangle}
\newcommand{\bra}{\langle}
\newcommand{\twist}{\mathcal{T}}
\newcommand{\C}{\mathbb{C}}

\title{Twist operators and pseudo %-Renyi
entropies\break in two-dimensional momentum space}
\author[1]{Giancarlo Camilo\footnote{Email: \texttt{gcamilo@iip.ufrn.br}}}
\author[1]{Andrea Prudenziati\footnote{Email: \texttt{andrea.prudenziati@gmail.com}}}
\affil[1]{\footnotesize International Institute of Physics, Federal University of Rio Grande do Norte\break Campus Universit\'ario, Lagoa Nova, 59078-970, Natal, RN, Brazil}
\date{\vspace{-5ex}}

\begin{document}

\maketitle

\begin{abstract}
We use a replica trick construction to propose a definition of branch-point twist operators in two dimensional momentum space and compute their two-point function. The result is then tentatively interpreted as a pseudo R\'enyi entropy for momentum modes.
\end{abstract}

%Note: there are commands \verb$\gian{}$ and \verb$\andrea{}$ to insert comments in separated colors.

\section{Introduction}\label{sec:intro}

Entanglement and R\'enyi entropies in Conformal Field Theories (CFTs) are an interesting and well studied subject. On one side, they provide an accurate measure for the entanglement of a pure state  \cite{nielsen00,Calabrese:2004eu}; on the other, stemming from the work of Ryu and Takayanagi \cite{Ryu:2006bv,Hubeny:2007xt,Dong:2016fnf}, they admit a remarkably simple holographic dual description, hinting at an intriguing relationship between entanglement and gravity/geometry that has been the subject of many works, e.g. \cite{Swingle:2009bg,VanRaamsdonk:2010pw,Maldacena:2013xja,Nishioka:2018khk}. 

Given that the study of conformal symmetry and its implications is usually done in real space, where the action of the conformal generators becomes a straightforward extension of the Poincaré transformations, it is no surprise that the studies of entanglement and R\'enyi entropies have so far been restricted mostly to degrees of freedom associated with spatial regions. In real space a well developed formalism \cite{Calabrese:2009qy} allows to transform R\'enyi entropies for two dimensional CFTs to correlation functions of branch-point twist operators. Being the latter local primary operators makes the computation analytically tractable in a number of cases  \cite{Calabrese:2009qy,Casini:2009sr,Caraglio:2008pk,Casini:2011kv,Calabrese:2009ez,Calabrese:2010he, Casini:2006hu,Bianchi:2015liz,Bueno:2015qya,CastroAlvaredo:2009ub};  a complementary approach from the dual geometric viewpoint usually provides simpler results in the classical large $G_N$ approximation \cite{Hartman:2013mia,Faulkner:2013yia,Bueno:2015xda}. 

Recently, motivated by applications in the context of holographic cosmology, there has been a surge of interest in the study of CFT correlators in momentum space \cite{Bzowski:2013sza,Bzowski:2019kwd,Bzowski:2020kfw,Isono:2018rrb,Gillioz:2019iye,Anand:2019lkt}. A related, natural question is how to make sense of R\'enyi and entanglement entropies between field modes in momentum space. In principle, this object may capture more general correlations that are intrinsically non-local from a position space point of view. However, the standard replica calculation of R\'enyi and von Neumann entanglement entropies in CFTs  has not yet been extended to momentum space, not to mention the absence of a holographic dual construction. As a result, very little is known about this object beyond the case of free quantum field theories where there is a clear factorization of the Hilbert space into modes with opposite momenta \cite{Balasubramanian:2011wt,Alves:2017fjk,Grignani:2016igg,PandoZayas:2014wsa,Brahma:2020zpk,Hsu:2012gk}. Our goal in this paper is to take a first step in this direction by introducing a notion of momentum space twist operators in $2d$ CFTs and calculating their two-point correlators. 

The definition follows closely the real space construction via the replica trick; in particular, it involves rotations in the full complex-$k$ plane. As we shall discuss, this fact implies that the method is not suited for the replica calculation of entanglement between regions of momentum space, since this would require fixed Euclidean time slices. Instead, it is more likely related to a notion of pseudo-entropy recently introduced in \cite{Nakata:2020fjg}.

The paper is organized as follows: in Section \Sec{xspace} we review the standard position space replica trick formalism and the definition of twist operators. In Section \Sec{kspace} we introduce our definition of twist operator in a two dimensional momentum plane and relate it with the Fourier transform of the position space twist operator. In Section \Sec{twopoint} we compute the two-point function of twist and anti-twist operators at arbitrary points in momentum space. This computation is carried on either using polar coordinates in Appendix \ref{poslsec} and spherical coordinates in Appendix \ref{possphsec}, in both cases restricting to points either only in the upper or lower half plane, the more complicated and somehow less relevant mixed case being discussed in Appendix \ref{polarnegl}. Finally, in Section \Sec{sp} we discuss the physical interpretation of our results in terms of pseudo entropies and some relevant physical properties. Section \Sec{conclusions} contains final remarks and a few open questions.

\section{Twist operators and the replica trick in position space}\label{sec:xspace}

Here we briefly review the standard replica trick construction of \cite{Calabrese:2009qy}. Let us consider the vacuum state $\rho_t=|0\ket\bra 0|_t$ of a quantum field theory defined on a Euclidean  manifold $\mathbb{R}_t\times\mathbb{M}$, where $\mathbb{M}$ is some $d$-dimensional spacelike manifold. Now fix a time slice $t\equiv0$, choose a spatial subregion $A\subset\mathbb{M}$, and then trace out the complementary region $A_c\equiv\mathbb{M}\setminus A$ to get the reduced state $\rho_A=\Tr_{A_c}\,\rho_0$. The corresponding entanglement entropy (EE) is defined as
\begin{equation}
S_A = -\Tr\,(\rho_A\log\rho_A)\,.
\end{equation}
This object involves the $\log$ of an infinite-dimensional operator and is typically hard to compute from scratch. 
The replica trick is based on the observation that the EE can be obtained indirectly by first calculating the R\'enyi entropies of integer order $n$, defined by
\begin{equation}\label{eq:Renyidef}
S^{(n)}_A= \Tr\,\rho_A^n\,,
\end{equation}
then analytically continuing the result for real $n$ such that
\begin{equation}\label{eq:EEdef}
S_A = -\partial_n S^{(n)}_A\big|_{n=1}\,.
\end{equation}
This has the clear advantage that  $\Tr\,\rho_A^n$ involves only integer powers of the reduced state $\rho_A$. 

The replica trick provides a practical way to calculate this object by considering a set of $n$ copies or {\it replicas} of the original space. Each $\rho_A$ can be represented as a path integral with a cut along the subregion $A$ at the time slice $t=0$ \footnote{Where the boundary configuration of fields along the two sides of the cut define the matrix element of $\rho_A$. }, and therefore $\Tr\,\rho_A^n$ is obtained by simply gluing a sequence of these $\rho_A$ along the cuts in a cyclic way. The resulting path integral is the partition function $\mathcal{Z}_n(A)$ on this $n$-fold cover $\mathcal{R}_n$ of the original spacetime or, to be more precise,
\begin{equation}\label{eq:TrZn}
\Tr\,\rho_A^n = \frac{\mathcal{Z}_n(A)}{\mathcal{Z}_1^n}\,,
\end{equation}
where $\mathcal{Z}_1$ (the partition function over an entire sheet having no cuts) ensures the unit trace of $\rho$ for each copy. 

In 2 dimensions \footnote{The following construction works also in higher dimensionality but for the twist operator definition that is obviously limited to the plane. Some discussion on the implementation of twist lines for three dimensional theories can be found in section 6.2 of \cite{Prudenziati:2019pln}} the partition function on this $n$-sheeted Riemann surface $\mathcal{R}_n$ can be calculated in certain simple cases where the surface has zero curvature everywhere except at a finite number of points. This is the case, for instance, where $A$ is an interval $[x_1,x_2]\subset\mathbb{R}$ of size $\ell=|x_1-x_2|$, in which the only points of non-vanishing curvature are the boundary points $x_1,x_2$ separating the regions $A$ and $A_c$. The partition function $\mathcal{Z}_n(A)$ in this case can be expressed as a path integral on a single sheet with identified cut $\mathcal{R}_1$, having implemented the $n$-sheeted structure of the Riemann surface by appropriate boundary conditions on n-copies of the fields inside region $A$, namely, 
\begin{align}\label{eq:ZnC}
\mathcal{Z}_n(A) &= \int [d\varphi]\,\exp\left[-\int_{{\mathcal R}_n} d\tau\,dx\,{\cal L}[\varphi]\right]\notag\\
&= \int_{\varphi_i(x,0^+)=\varphi_{i+1}(x,0^-)\,\forall\, x\in A} \,[d\varphi_1 \cdots d\varphi_n]\,\exp\left[-\int_{\mathcal{R}_1}d\tau\,dx\,{\cal L}^{(n)}[\varphi_1,\ldots,\varphi_n]\right]\,,
\end{align}
where the Lagrangian ${\cal L}^{(n)}[\varphi_1,\ldots,\varphi_n]={\cal L}[\varphi_1]+\ldots+{\cal L}[\varphi_n]$ due to locality (i.e., it does not depend explicitly on the Riemann surface). By introducing the twist fields $\twist_n(x_1)$ and $\bar{\twist}_n(x_2)$ that generate the branch cut along $A$ associated with the cyclic permutation symmetries $i\to i+1$ and $i+1\to i$ between the replicas, respectively, the partition function \eq{ZnC} can be expressed as a 2-point correlator between the twist fields,
\begin{align}\label{eq:Zntwist}
\mathcal{Z}_n(A) &= \int [d\varphi_1 \cdots d\varphi_n]\,\exp\left[-\int_{\mathcal{R}_1}d\tau\,dx\,\left({\cal L}[\varphi_1]+\ldots+{\cal L}[\varphi_n]\right)\right]\,\twist_n(x_1)\,\tilde{\twist}_n(x_2)\notag\\
&= \mathcal{Z}_1^n\,\bra \twist_n(x_1)\,\bar{\twist}_n(x_2) \ket_{\mathcal{R}_1}\,.
\end{align}
Here we have assumed that the definition of $\twist_n,\bar{\twist}_n$ is such that the righthand side is dimensionless, and $\mathcal{Z}_1^n$ denotes the path integral with no insertions (i.e., the $n$-th power of the standard partition function). 

With the help of \eq{Zntwist}, the R\'enyi entropy \eq{Renyidef} becomes
\begin{equation}\label{eq:Renyitwist}
S^{(n)}_A= \bra \twist_n(x_1)\,\bar{\twist}_n(x_2) \ket_{\mathcal{R}_1}\,.
\end{equation}
The task now has been reduced to calculating the correlator between the two twist operators. When the theory at hand is a CFT and $\mathcal{R}_1$ the complex plane, the result is well-known \cite{Calabrese:2009qy} 
\begin{equation}\label{eq:correlatortwist}
\bra \twist_n(x_1)\,\bar{\twist}_n(x_2) \ket_{\C} = \mathcal{N}_n\,|x_1-x_2|^{-2\Delta_\twist}\,,\qquad \Delta_\twist = \frac{c}{12}\left(n-\frac{1}{n}\right)\,,  
\end{equation}
since $\twist_n$ is a primary operator of dimension $\Delta_\twist$. Here $\mathcal{N}_n$ is a normalization constant that we shall parametrize as $\mathcal{N}_n\equiv \alpha_n\,a^{2\Delta_\twist}$ for some small length scale $a$ and dimensionless constant $\alpha_n$ (notice that $\alpha_1=1$ since $\Tr\,\rho_A=1$). Plugging in (\ref{eq:correlatortwist}) into (\ref{eq:Renyitwist}) and using (\ref{eq:EEdef}) one immediately gets the celebrated result for the EE
\begin{equation}\label{eq:EExspace}
S_A = \frac{c}{3}\log\frac{\ell}{a}+\text{const}\,.
\end{equation}
The first term gives the area law and has the universal coefficient $c/3$, while the non-universal constant contribution is reminiscent of the normalization constant $\alpha_n$, that is, $\text{const} = \lim_{n\to1}\,(1-n)^{-1}\log\alpha_n$.

\section{Momentum space twist operators}\label{sec:kspace}

The idea here is to generalize the construction of the previous section to momentum space. In particular, our goal will be to compute a correlator of momentum space twist operators i.e.
\begin{equation}\label{eq:Renyitwistk}
S^{(n)}_A= \bra \twist_n(k_1)\,\bar{\twist}_n(k_2) \ket_{\C}\,.
\end{equation}
For simplicity we focus on the two dimensional complex plane in momentum coordinates $k=k_R+i k_I$ and $\bar{k}=k_R-i k_I$. 
Here we keep the insertion points $k_1,k_2$ completely arbitrary; %in position space an analogous computation could have been made as well still producing (\ref{eq:correlatortwist}) but, with $x_1,x_2$ not sitting at the same time $t$ (or on any boosted space-like interval) it would make the interpretation of the replica path integral a bit more obscure; o
On the momentum plane the additional difficulty is in not having a time direction at all, and therefore \eq{Renyitwistk} should not be understood in general as a Rényi entropy (i.e., it does not prepare the $\Tr\rho_A^n$ for any quantum state $\rho_A$). A discussion on the physical interpretation of this correlator  will follow later. 

The first question that comes to mind is how to even make sense of the momentum space twist operators $\twist_n(k)$ appearing above. 
What we do know about the twist operators, either in real or momentum space, is that by definition they produce a branch cut in the space which, when crossed, yields a jump by one in the replica index of the local fields, as explained before. So let us pick a twist field $\twist_n(k_1)$ and some probe operator $\Op_i(k)$. If $k$ circles around $k_1$ by an angle $\delta$ in, say, the anticlockwise direction, after coming back to the original position ($\delta=2\pi$) the operator will have jumped in its replica index, without any other effect, regardless of where the branch cut was located. Schematically,
\begin{equation}\label{twist}
   \twist_n(k_1)\,\Op_i(k)\quad\xrightarrow[]{\delta\text{ rot.}}\quad \twist_n(k_1)\,\Op_i(\tilde{k}(k_1,\delta)) \quad\xrightarrow[]{\delta=2\pi}\quad \twist_n(k_1)\,\Op_{i+1}(k)\,,
\end{equation}
where for simplicity we define the rotating momentum $\tilde{k}(k_1,\delta)\equiv k_1+(k-k_1)e^{i\delta}$ (and we have assumed $\delta$ after the first arrow small enough to not have crossed the branch cut location). 
In order to make contact with our knowledge of position space twist operators, let us Fourier transform the above expression by assuming that the twist operator at $k_1$ in momentum space is represented, in real space, as some distribution $\sigma_n$ that will depend on $k_1$ and possibly also on the position $x$ and the momentum of the rotating operator $k$. 
That is, before and after the rotation by $\delta$ we have, respectively,
\begin{align}\label{fourier}
    \twist_n(k_1)\Op_i(k) &= \frac{1}{2\pi}\int \frac{d^2 x}{2}\; e^{\frac{i}{2}(\bar{k}x+k\bar{x})}\sigma_n\,\Op_i(x)\notag\\ 
    \twist_n(k_1)\Op_i(\tilde{k}(k_1,\delta)) &= \frac{1}{2\pi}\int \frac{d^2 x}{2}\; e^{\frac{i}{2}(\bar{\tilde{k}} x+\tilde{k}\bar{x})}\sigma_n\,\Op_i(x)\,.
\end{align}
It is important to note that this expression is at a classical level and the pair $\twist_n(k_1)\Op_i(k)$  should be understood as a function $\Op_i(k)$ in presence of a momentum twist operator such that  (\ref{twist}) holds; therefore $\twist_n(k_1)\Op_i(k)$ is not, at this level, a two-point function. Analogously $\sigma_n\Op_i(x)$ represents the Fourier transform of $\Op_i(k)$ in presence of some (unknown) distribution whose job is to keep (\ref{twist}) valid on the right hand side of the equation.
We have also implicitly imposed that each  replica index value has its own Fourier transform and indicated $x=x_R+i x_I$,  $\bar{x}=x_R-i x_I$. By changing the integration variable to 
\begin{equation}
\tilde{x}(k_1,k,x,\delta)\equiv \frac{\bar{\tilde{k}}}{\bar{k}}x=x\left[ \frac{\bar{k}_1}{\bar{k}}+\left(1-\frac{\bar{k}_1}{\bar{k}}\right)e^{-i\delta}\right].
\end{equation}
the last expression can be rewritten as 
\begin{equation}\label{fourier2}
  \twist_n(k_1)\Op_i(\tilde{k}(\delta,k_1))
  = \frac{1}{2\pi}\int \frac{d^2 \tilde{x}}{2|\frac{\tilde{k}}{k}|^2}\; e^{\frac{i}{2} \bar{k} \tilde{x}+k\bar{\tilde{x}}}\sigma_n\,\Op_i\left(\frac{\bar{k}}{\bar{\tilde{k}}}\tilde{x}\right)\,.
\end{equation}
In the $\tilde{x}$-coordinate system the rotation around $k_1$ in the momentum plane has been transformed to the path parameterized by $\delta$ in real space, followed by the operator
\[
\Op_i\left(\frac{\tilde{x}}{ \frac{\bar{k}_1}{\bar{k}}+(1-\frac{\bar{k}_1}{\bar{k}})e^{-i\delta}}\right).
\]
One can show that this is a circle as well, whose center is located at
\begin{equation}\label{eq:ec}
    x_1(k_1,k,\tilde{x})\equiv \frac{\frac{k_1}{k}}{\frac{k_1}{k}+\frac{\bar{k}_1}{\bar{k}}-1}\,\tilde{x}.
\end{equation}

For consistency, after a full rotation by $2\pi$ the only effect on the real space operator must be an increment of one in its replica index. 
This constrains the otherwise mysterious distribution $\sigma_n$ to be a real space twist operator located in $x_1$ times a proportionality factor that may depend nontrivially on the phase $\delta$ and the momenta $k_1,k$ and is only restricted by the boundary conditions at $\delta=0$ and $2\pi$, that is,
\begin{equation}\label{sigma}
    \sigma_n(k_1,k,\tilde{x})=\twist_n(x_1)\,\rho(k_1,k,\delta) \,\qquad \rho(k_1,k,0)=\rho(k_1,k,2\pi)=1.
\end{equation}
The factor $\rho(k_1,k,\delta)$ may be used to absorb the Jacobian in (\ref{fourier2}) but will otherwise play no essential role in our discussion. 

Now let us pick $\delta=0$, so that $\tilde{x}=x$; moreover we rewrite  either (\ref{fourier2}) or (\ref{fourier}) as \footnote{The delta function here is normalized to give unity when integrated over real and imaginary parts of $k$. }
\begin{align}
    \twist_n(k_1)\Op_i(k)&=\frac{1}{2\pi}\int \frac{d^2k_0}{2} \twist_n(k_1)\,2\pi \delta^2(k-k_0)\Op_i(k_0)\notag\\
    &= \frac{1}{(2\pi)^2}\int \frac{d^2k_0}{2} \int \frac{d^2 x}{2}\; e^{\frac{i}{2}((\bar{k}-\bar{k}_0)x+(k-k_0)\bar{x})}\sigma_n(k_1,k-k_0,x) \Op_i(k_0) \notag \\
    &=\frac{1}{(2\pi)^2}\int \frac{d^2k'}{2} \int \frac{d^2 x}{2}\; e^{\frac{i}{2}(\bar{k}'x+k'\bar{x})}\sigma_n(k_1,k',x) \Op_i(k-k') \,.
\end{align}
This and (\ref{sigma}) show that the right hand side is a convolution between the composite function $\sigma_n(x_1(k'))$ and the operator.
To isolate the twist operator alone we now choose $\Op_i(k)=1$, then using (\ref{sigma}) we obtain
\begin{equation}\label{eq:main}
    \twist_n(k_1) = \frac{1}{(2\pi)^2}\int \frac{d^2k}{2} \int \frac{d^2 x}{2}\; e^{\frac{i}{2}(\bar{k}x+k\bar{x})}\,\twist_n(x_1).
\end{equation}
This is the main relation we will use to transform momentum space correlation functions of twist operators into their integrated real space counterparts. Notice that, despite of the Fourier-like appearance, the relation between momentum- and real-space twist operators is more complicated than a simple Fourier transform due to the intricate structure of $x_1$ shown in \eq{ec} together with the extra integration over $k$. 

It is instructive to express \eq{main} in terms of the standard Fourier transform of the position-space twist operator, 
\begin{equation}
\tilde{\twist}_n(p)\equiv \frac{1}{2\pi}\int \frac{d^2 x}{2} e^{\frac{i}{2}(\bar{p}x+p\bar{x})}\,\twist_n(x)    
\end{equation}
which itself does \textbf{not} have the properties of a twist operator in momentum-space and hence is denoted with a tilde here to avoid confusion with our $\twist_n(k)$. This can be done by changing the integration variable $x$ in \eq{main} to $x_1=x\,c_1(k_1,k)$ where
\begin{equation}
c_1(k_1,k) = \frac{\frac{k_1}{k}}{\frac{k_1}{k}+\frac{\bar{k_1}}{\bar{k}}-1}=\frac{k_1\bar{k}}{2\,\text{Re}(k_1\bar{k})-|k|^2}\,,
\end{equation}
which introduces a Jacobian $|c_1|^{-2}$. After some simplifications we end up with 
\begin{align}
\twist_n(k_1)
&= \frac{1}{2\pi|k_1|^2}\int \frac{ d^2k}{2|k|^2}\,\big(2\,\text{Re}(k_1\bar{k})-|k|^2\big)^2\, \tilde{\twist}_n(p_1)\,,\qquad p_1=\frac{k}{\bar{c}_1}=\frac{2\,\text{Re}(k_1\bar{k})-|k|^2}{|k_1|^2}k_1\,.
\end{align}
% with 
% \begin{align}
% p_1=\frac{k}{\bar{c}_1}=\frac{2\,\text{Re}(k_1\bar{k})-|k|^2}{|k_1|^2}k_1\,.
% \end{align}
This begs for the further substitution $k\to k'\equiv \frac{1}{2|k_1|^2}\bar{k}_1k=\frac{k}{2k_1}$, which gives 
\begin{align}
\twist_n(k_1) %&= %\frac{1}{2\pi|k_1|^2}\,\int \frac{ d^2k'}{2|k'|^2}\,\big(4|k_1|^2\,\text{Re}(\bar{k'})-4|k_1|^2\,|k'|^2\big)^2\,\tilde{\twist}_n(p_1)\notag\\
&= \frac{16|k_1|^2}{2\pi}\,\int \frac{ d^2k}{2|k|^2}\,\big(k^R-|k|^2\big)^2\, \tilde{\twist}_n(p_1)
\end{align}
where now 
\begin{align}\label{eq:p1def}
p_1=4\big(k^R-|k|^2\big)k_1
\end{align}
and we use the shorthand notation $z^R\equiv\text{Re}(z),z^I\equiv\text{Im}(z)$.
We can then use the reality of $k^R-|k^2|$ to write $\big(k^R-|k^2|\big)^2%=\big(p_1/4k_1\big)^2=\big|p_1/4k_1\big|^2
=|p_1|^2/16|k_1|^2$ and arrive at the simple expression 
\begin{align}\label{eq:main2}
\twist_n(k_1) &= \frac{1}{2\pi}\int \frac{ d^2k}{2|k|^2}\,|p_1|^2\,\tilde{\twist}_n(p_1) \,.
\end{align}
Once again this emphasizes the fact that the momentum space twist operator is more complicated than a simple Fourier transform of the real space one. In what follows we will make use this form to compute the two-point function of $\twist_n(k)$.

\section{Two-point function}\label{sec:twopoint}

Let us now consider the two point function of momentum-space twist operators located at $k_1,k_2$. Our proposal is to promote \eq{main2} to a quantum expression acting on each of the twist operators inside an $n$-point function. For the simplest case of a two point function we are then led to the equation
\begin{align}\label{twistk}
\big\bra\twist_n(k_1)\bar{\twist}_n(k_2)\big\ket &= \frac{1}{(2\pi)^2}\int \frac{d^2k\,d^2k'}{4|k|^2|k'|^2}\,|p_1|^2\,|p_2|^2\,\big\bra\tilde{\twist}_n(p_1)\tilde{\bar\twist}_n(p_2)\big\ket\,,
\end{align}
where $p_1=p_1(k_1,k)$ is given in \eq{p1def} and $p_2=p_2(k_2,k')$ is defined analogously. The two-point function of tilded quantities appearing on the right is nothing but the Fourier transform of the real-space two point correlator \eq{correlatortwist}, which takes the form (here $\Delta\equiv\Delta_\twist$) 
\begin{align}\label{twistp}
\big\bra\tilde{\twist}_n(p_1)\tilde{\bar\twist}_n(p_2)\big\ket &= (2\pi)^2\frac{\mathcal{N}_n\pi \,2^{2(1-\Delta)}\Gamma(1-\Delta)}{\Gamma(\Delta)}\,\delta(p_1+p_2)\,|p_1|^{2(\Delta-1)}\,.
\end{align}
We are then left with 
\begin{align}\label{eq:Renyi}
\big\bra\twist_n(k_1)\bar{\twist}_n(k_2)\big\ket &= \frac{\mathcal{N}_n\pi \,2^{2(1-\Delta)}\Gamma(1-\Delta)}{\Gamma(\Delta)}\int\frac{d^2k\,d^2k'}{4|k|^2|k'|^2}\,\delta(p_1+p_2)\,|p_1|^{2(\Delta+1)}\,.
\end{align}

The double integral $I=I(k_1,k_2;\Delta)$ above is non-vanishing only when the two momenta $k_1,k_2$ are colinear, i.e., when $k_2=\lambda k_1$ with $\lambda\in\mathbb{R}$. It is computed in detail, in the limit of $\Delta\rightarrow 0$, in two different ways in Appendices \ref{poslsec} and \ref{possphsec} for $\lambda>0$  and in Appendix \ref{polarnegl} for $\lambda<0$. The two-point function can then be expressed in a fairly simple way by isolating the $\lambda$ dependence  as
\begin{equation}
   \big\bra \twist_n(k_1)\bar{\twist}_n(k_2)\big\ket = \frac{\mathcal{N}_n\pi \,2^{2(1-\Delta)}\Gamma(1-\Delta)}{\Gamma(\Delta)}\,  \delta(0)\,|k_1|^{2(\Delta+1)}4^{2\Delta+1} F(\lambda,\Delta)\,.
\end{equation}
An explicit regularization scheme for the $\delta(0)$ divergence is also discussed in Appendix \ref{poslsec} and takes the form $\delta(0)=\lim_{\epsilon^{R/I}\to 0}\delta(k_1^R\epsilon^I-k_1^I\epsilon^R)$ \footnote{For simplicity we will continue writing $\delta(0)$ unless where the explicit form becomes relevant.}.

In analogy with the standard, position-space derivation of entanglement entropy from the two point function of twist operators, we would like to define the function 
\begin{equation}\label{ententdef}
S(k_1,k_2)=-\partial_n \bra\twist_n(k_1)\bar{\twist}_n(k_2)\ket_{n=1}\,.
\end{equation}
As discussed in the next Section, this object should be viewed as a pseudo entanglement entropy in the sense of \cite{Nakata:2020fjg} instead of a standard von Neumann entanglement entropy. Using
\[
\partial_n \frac{\Gamma(1-\Delta(n))}{\Gamma(\Delta(n))}\Big|_{n=1}=\frac{c}{6}, \;\;\;\; \frac{\Gamma(1-\Delta(n))}{\Gamma(\Delta(n))}\Big|_{n=1}=0, \;\;\;\; \mathcal{N}_1=1
\]
one gets 
\begin{equation}\label{ententropy}
    S(k_1,k_2)=-c\,\frac{8}{3} \;\pi \;\delta(0)\,  |k_1|^2 F(\lambda,0)\;.
\end{equation}
%This is the main reason why we are interested in the limit $\Delta\rightarrow 0$ of \eq{Renyi}.

For the positive $\lambda$ case one can use (\ref{poslamb}) (or (\ref{ilo})) such that 
\begin{equation}\label{posf}
    F(\lambda,0)= \frac{\pi^2}{4}(\lambda+1)
\end{equation}
to get the simple-looking result
\begin{equation}
    S(k_1,\lambda k_1)_{\lambda>0}=-c\,\frac{2}{3} (1+\lambda)\;\pi^3 \, \delta(0)\,|k_1|^2 \;.
\end{equation}

For negative $\lambda$ the situation is less clear since the result contains extra divergences, so we leave the explicit results to Appendix \ref{polarnegl}. Some discussion on its physical significance is in the next section.

\section{Physical interpretation}\label{sec:sp}

\subsection{Pseudo-Renyi entropy and Twist operator locations}

In  position-space, %and in time-independent theories
and for time-independent configurations (such that Wick rotation to Euclidean signature is available), by setting the two twist operators at a constant time slice their two point function computes the R\'enyi entropy for modes living on the spatial interval in between. This interpretation has already been discussed and is based on the path integral formalism of quantum field theory. If we Fourier transform only along the spatial direction while keeping the time coordinate, the situation is analogous and a twist operator two point function in this situation at positions $(k_1,t)$ and $(k_2,t)$ would compute the R\'enyi entropy at time $t$ for modes contained between momenta $k_1, k_2$ and their complement Hilbert space. When the Fourier transform is taken over both directions, however, the physical interpretation becomes less clear. What we have provided here is a computation of the two point function of twist operators in the two dimensional momentum space at arbitrary positions $(k^0_1, k^1_1)$ and $(k^0_2, k^1_2)$, and the interpretation we will tentatively explore is via a generalization of R\'enyi entropy along the lines of the recently introduced pseudo R\'enyi entropy \cite{Nakata:2020fjg}.%, for modes contained on the interval connecting the points $k_1, k_2$. 

%To make this interpretation a little clearer 
The first point we would like to clarify is that path integrating on the momentum plane up to a certain energy $k^0$ (or along some spacelike line at each point in the energy-momentum variables) produces some quantum state in the Hilbert space in the same way as the usual spacetime path integration from $t=-\infty$ to $t_0$ (or along some spacelike line in position-space). 

Path integral in quantum mechanics with momentum boundary conditions and the usual time direction had been already considered by Feynmann in the appendix B of his paper \cite{Feynman:1951gn} (see also \cite{Klauder_2003} ), by Fourier transforming the end-points coordinates. In quantum field theory the situation is even simpler; what we have to do is simply to pick as canonical variables in the integral the fields in momentum space $\phi(k,t)$ instead than in position space $\phi(x,t)$. The complicated part then is to interpret a path integral where the time direction has been replaced by energy.

Let us then consider the formal expression for such a path integral:
\begin{equation}\label{ni}
    \int\mathcal{D}\phi(k) e^{-\tilde{S}(\phi(k))}\,,
\end{equation}
where $\mathcal{D}\phi(k)$ is the integration measure on field configurations at each point $k$ and $\tilde{S}(\phi(k))$ is the action for the fields that have been Fourier transformed; for instance a massless scalar field of position space Lagrangian density $L(\phi(x))=\frac{1}{2}\partial_{\mu}\phi(x)\partial^{\mu}\phi(x)$ is described in momentum space by an action $\tilde{S}(\phi(k))=\frac{1}{2(2\pi)^2}\int d^2k k^2\phi(k)\phi(-k) $. This path integral will also include some boundary conditions that we are going to discuss later on.
Now we can Fourier transform this field; for simplicity we only Fourier transform explicitly the time direction and keep either space or momenta in the other, as we have seen either choice leads to the same physical interpretation by picking the correct canonical variables. Then with $t=$ time and $k=$ energy:
\begin{equation}\label{msp}
    \int\mathcal{D}\phi(k) e^{-\tilde{S}(\phi(k))}=\int\mathcal{D}\left[\int dt e^{i k t}\phi(k)\right] e^{-S(\phi(t))}\,.
\end{equation}
The integration measure should be interpreted as the UV limit of some lattice regularization, where $k$ picks all the energy values $k_0,\dots, k_n$ on the chosen domain, starting from $k_0$ at the boundary value $k_i$ and ending with $k_n=k_f$. By Fourier transforming however, $t$ ranges from $-\infty$ to $+\infty$:
\begin{align}\label{lc}
    \int\mathcal{D}\phi(k) &\sim \int d\phi(k_0)d\phi(k_1)d\phi(k_2)\dots d\phi(k_n)\Big|_{k_i=k_0<k_1<k_2\dots<k_n=k_f} \notag\\
    &=\int_{-\infty}^{\infty} dt_0 dt_1 dt_2\dots dt_n e^{i(k_0t_0+k_1t_1+k_2t_2+\dots+k_n t_n)}\int d\phi(t_0)d\phi(t_1)d\phi(t_2)\dots d\phi(t_n)\Big|_{k_i=k_0<k_1<k_2\dots<k_n=k_f} 
\end{align}
The time variables are not ordered but, since the fields $\phi(t)$ are classical variables in the path integral, their permutation presents no problem. Suppose for instance that we are integrating over some region where $t_1<\dots< t_0<\dots <t_2<\dots$. Then the integration measure can be reordered as $d\phi(t_1)\dots d\phi(t_0)\dots d\phi(t_2)\dots$, or the variables $t_i$ renamed accordingly. In the following we then assume that $t_i\leq t_{i+1}$ always. 

The values of the set of $\{t_i\}$, that we eventually want to interpret as a set of lattice points as we did for the $\{k_i\}$, change at each point in the integration domain $\mathcal{I}$ over $t_0,\dots t_n$. So the lattice in the time direction will not cover some fixed interval as defined with the energies, but some over constantly changing region $\mathcal{C}_t$. We define such region $\mathcal{C}_t$ at each point in $\mathcal{I}$ by adding a small neighborhood of size $\epsilon_t$ around any point $t_i$, and then joining all these closed sets. In this way the resulting path integral will be interpreted to cover the time region in which the values of the $\{t_i\}$ are dense in a lattice sense. Obviously $\mathcal{C}_t$ depends continuously on the point in $\mathcal{I}$ at which we are. 

The right hand side of (\ref{lc}) will then become the integration measure of some path integral on the time direction, spanning a time set $\mathcal{C}_t$ that may be quite complicated. If  $\mathcal{C}_t$ is connected, which happens if $t_{i+1}-t_i\leq\epsilon_t$ for any $i$, then the path integral interpretation is clear: it produces some state at $t_0$ by the field configuration at that time, and then evolves it continuously until $t_n$, with the caveat that in this case the boundary conditions are not fixed but integrated over. The path integral in this case will then compute the scalar product between an initial state at $t_0$, $\bra \phi_{t_0} |$ and a final state at the boundary time $t_n$, $\bra \phi_{t_0}e^{-(t_n-t_0)H} |\phi_{t_n} \ket$, integrated over all the possible $\bra \phi_{t_0} |$, $\bra \phi_{t_n} |$. If the integration domain is disconnected instead, we can interpret the path integral as the product of as many path integrals as the connected components of $\mathcal{C}_t$. On each connected interval we will have a path integral computing the scalar product between the corresponding states produced by the boundary conditions at the end-points, so for each connected interval a complex number $|z|\leq 1$ is produced. The last interval will determine the final state evolved up to the last value of $t_{n-1}$, before taking the final scalar product at $t_n$. So in any case the path integral  (\ref{msp}) ends up computing an infinite linear combination of states, as in (\ref{lc}), in the Hilbert space of the theory, produced by standard  path integration over the time direction, which is again a state in the Hilbert space (up to normalization issues that should be solved by appropriate definition of (\ref{ni})).

However, contrary to the usual time-independent position-space case with vanishing boundary conditions, where the path integration from $t=-\infty$ to the cut between twist operators at $t=t_0$ produces some ket state\footnote{The vacuum state can be replaced by any other state by changing boundary conditions at $t=\pm\infty.$} $|0\ket_{t_0}$ while the one from $t_0$ to $t=\infty$ produces its bra version  $\bra 0|_{t_0}$, here the ket and bra states will be in general different. This is the reason why we are pointing towards a pseudo-R\'enyi interpretation where the usual reduced density matrix $\rho_A=\Tr_B |\psi\ket\bra \psi|$ appearing in the calculation of the entropy is replaced by a \emph{reduced transition matrix}
\begin{equation}
T_A^{\psi|\phi}=\Tr_B \frac{|\psi\ket\bra \phi|}{\bra\psi|\phi\ket}
\end{equation}
depending on a pair of different states. 
Once this is settled the replica trick construction should parallel the one in position-space. First we would fix field boundary conditions along the two sides of the cut to identify the matrix element of, in this case, the transition matrix $T_A^{\psi|\phi}$, and then gluing successively $n$-sheets in order to calculate the $\Tr[(T_A^{\psi|\phi})^n]$. Finally the one-sheeted version with $n$-field copies is implemented by introducing our notion of momentum space twist operators $\twist(k)$.

Analogously to the position-space analysis the field boundary configuration on the two sides of the cut produce the matrix elements for the transition matrix (or the reduced density matrix for Renyi entropies) and should therefore belong to the Hilbert space of the theory under consideration. Two novel issues then appear in comparison with the position space-analysis: the first difference is that physical modes here must respect the mass shell constraint, which limits the available phase space; momentum for a massless field should for instance be placed on the light-cone while in an interacting theory the energies are shifted below or above \footnote{ In particular, if the interaction is strong enough the total energy of a field mode may be even negative}. The second difference comes from the fact that our result is non-vanishing only when the two momenta at the end points of the cut are proportional, i.e., $k_2=\lambda k_1$ with $\lambda\in\mathbb{R}$. 

Joining these two observations we are led to the conclusion that the two point function of twist operators are quite constrained and the two twist operators should then be placed generically anywhere on a straight line passing through the origin and should additionally belong to the positive light cone for a free theory, or contained in its interior for a repulsive interaction or anywhere outside (including the lower half plane) for the attractive case:
\begin{itemize}
    \item Free theories: $k_1=(k_1^0, k_1^1)$, $k_2=(\lambda k_1^0, \lambda k_1^1))$ with $k_1^0=|k_1^1|$ and $\lambda>0$.
     \item Repulsive interaction: $k_1=(k_1^0, k_1^1)$, $k_2=(\lambda k_1^0, \lambda k_1^1))$ with $k_1^0\geq|k_1^1|$ and $\lambda>0$.
    \item Attractive interaction:  $k_1=(k_1^0, k_1^1)$, $k_2=(\lambda k_1^0, \lambda k_1^1))$ with $k_1^0\leq|k_1^1|$.
\end{itemize}
This in particular means that the negative $\lambda$ case is excluded but for attractive theories.

\subsection{Discussion on the sign}

Since the eigenvalues $\lambda_i$ of a reduced density matrix are the probabilities of the system to be in the corresponding eigenstate, by definition we have $1\geq\lambda_i\geq 0$ (and $\sum_i \lambda_i=1$), so in particular $1\geq\Tr \rho^n \geq 0$  for $n\geq 1$ and $\partial_n \Tr \rho^n\leq 0$. The obvious consequence is that $S=-\partial_n \Tr \rho^n|_{n=1}$ is always non-negative. By computing $\Tr \rho^n\propto \big\bra\twist_n(x_1)\bar{\twist}_n(x_2)\big\ket$ in real space and Fourier transforming this result in terms of $\big\bra\tilde{\twist}_n(p_1)\tilde{\bar\twist}_n(p_2)\big\ket$, we immediately see that the latter no longer has the required sign properties. 
This happens because (\ref{twistp}) contains a ratio of Gamma functions that behaves as $\partial_n \frac{\Gamma(1-\Delta(n))}{\Gamma(\Delta(n))}\Big|_{n=1}=\frac{c}{6}$ and $\frac{\Gamma(1-\Delta(n))}{\Gamma(\Delta(n))}\Big|_{n=1}=0$, so the derivative $-\partial_n \big\bra\tilde{\twist}_n(p_1)\tilde{\bar\twist}_n(p_2)\big\ket|_{n=1}\leq 0$ given that $c\geq 1$. But that is fine, since we are not attempting a probabilistic interpretation for $\big\bra\tilde{\twist}_n(p_1)\tilde{\bar\twist}_n(p_2)\big\ket$. %%So far no problems as we do not have any probabilistic interpretation for $\big\bra\tilde{\twist}_n(p_1)\tilde{\bar\twist}_n(p_2)\big\ket$, but we would eventually like to have it for (\ref{twistk}) that maintains its same exact sign properties. 

In fact, from the point of view of the pseudo entropies introduced \cite{Nakata:2020fjg} this is even expected. By construction these quantities are complex-valued since generic transition matrices are not even Hermitian, not to mention positive-definite operators. As shown by the authors, only in very special cases (such as particular qubit configurations or holographic states with a classical gravity dual) they conspire to produce real values. We notice that the Fourier transform (\ref{twistp}) grows monotonically with $n$ around $n=1$ %, negative on the left and positive on the right, 
for any value of the momentum $p_1$, and this is then inherited by (\ref{twistk}). As a result, the pseudo entropy $S$ which involves $-\partial_n$ must indeed be negative-definite. This result poses interesting questions on the possible interpretation of our two-point correlator and on the function (\ref{ententropy}). In particular, if some pseudo entropy reading is possible for the former it should belong to some very special subclass of the most general case leading to real-valued and negative entropy; for the latter instead we cannot avoid the speculation of whether simply changing the sign in the definition of (\ref{ententropy}) so that the result is  positive definite would lead to a well-defined entanglement measure.

\subsection{Symmetry condition for Entanglement Entropy under the exchange $k_1\leftrightarrow k_2$}

We would like to study in this section and the next the behaviour of the two point function with respect to exchange of the twist operator insertion points $k_1 \leftrightarrow k_2$, as well as rotation around the origin. 

The exchange  $k_1 \leftrightarrow k_2$ translates into a very specific condition for the function $F_{\pm}(\lambda,0)$ appearing in (\ref{ententropy}). We note that $(k_1,\lambda k_1)\rightarrow (\lambda k_1,k_1)$ can be achieved by the two subsequent transformations 
\begin{align}\label{tran}
   &\lambda \rightarrow \frac{1}{\lambda} 
   &&k_1 \rightarrow \lambda k_1 \,,
\end{align}
meaning that the pseudo entropy (\ref{ententropy}) transforms as
\begin{align}
 S(k_1,\lambda k_1)&=     -c\frac{8}{3} \;\pi \;\lim_{\epsilon^{R/I}\to 0}\delta(k_1^R\epsilon^I-k_1^I\epsilon^R)\,  |k_1|^2 F_{\pm}(\lambda,0) \rightarrow -c\frac{8}{3} \;\pi \;\lim_{\epsilon^{R/I}\to 0}\delta(\lambda k_1^R\epsilon^I-\lambda k_1^I\epsilon^R)\,  |\lambda k_1|^2 F_{\pm}(\frac{1}{\lambda},0)\notag \\
 &=-c\frac{8}{3} \;\pi \;\lim_{\epsilon^{R/I}\to 0}\delta(k_1^R\epsilon^I-k_1^I\epsilon^R)\,  |k_1|^2 |\lambda|F_{\pm}(\frac{1}{\lambda},0)\,.
\end{align}
It follows that invariance under $k_1 \leftrightarrow k_2$ implies that the function $F(\lambda,0)$ must have the symmetry\footnote{The modulus for $\lambda$ is important when $\lambda<0$ for which the condition applies as well.}
\begin{equation}\label{cond}
F(\lambda,0)=|\lambda|F(\frac{1}{\lambda},0)\,.
\end{equation}
We can immediately verify that our result (\ref{posf}) satisfies this condition.

Let us now show this symmetry explicitly by considering (\ref{eq:Ipolar}) equivalently written as
\begin{align}\label{eq:Ipolar2}
I = \int drd\theta dr'd\theta'\,\frac{1}{rr'}\,\delta(p_1+p_2)\,|p_2|^{2(\Delta+1)}\,,
\end{align}
and solving the delta argument (\ref{quadeqi}) in $r$ rather than in $r'$ by using (\ref{solr}). This solution is valid for $\lambda=0$ and can then be used to compute $F(1/\lambda)$ even for $\lambda\rightarrow\infty$. The new form of (\ref{integ}) becomes
\begin{align}\label{integnew}
I &= \frac{\delta(0)\,(4|\lambda k_1|)^{2(\Delta+1)}}{4}\int drd\theta dr'd\theta'\,\frac{(r'\cos\theta'-r'^2)^{2(\Delta+1)}}{r'T(r',\theta',\theta)} \frac{1}{r}\sum_{i=\pm} \delta(r-r_{*,i})\,.
\end{align}
We now rename variables to $r,\theta \leftrightarrow r',\theta'$ to get
\begin{align}\label{integnewc}
I &= \frac{\delta(0)\,(4|\lambda k_1|)^{2(\Delta+1)}}{4}\int drd\theta dr'd\theta'\,\frac{(r\cos\theta-r^2)^{2(\Delta+1)}}{rT(r,\theta,\theta')} \frac{1}{r'}\sum_{i=\pm} \delta(r'-r'_{*,i})\,.
\end{align}
where $r'_{*,i}$ is (\ref{solr}) with the variables renamed, in the same form of (\ref{solrp}) but with $1/\lambda\to\lambda$. The rest of the integral (\ref{integnewc}) is also the same of (\ref{integ}) but having replaced the position of $\lambda$ in the prefactor. So we conclude that, at $\Delta=0$, $I(\lambda)=|\lambda| I(1/\lambda)$, with the caveat that $I(\lambda)$ is valid for $\lambda\in \mathbb{R}/_{0}$ and
$I(1/\lambda)$ is valid for $\lambda\in \mathbb{R}/_{\infty}$. Extracting the function $F(\lambda)$ from the two cases we obtain exactly (\ref{cond}).

\subsection{Behaviour under momentum rotation $k_1\rightarrow e^{i\theta}k_1$}
In two Euclidean dimensions the time direction can be chosen to be whatever we want. Extracting the entanglement entropy from a two-point function of twist operators automatically assumes that this direction has been chosen either as orthogonal to the line connecting the two operators or somewhere in the light cone centered there, since the entanglement entropy should be interpreted as an entanglement measure either at fixed time or on a boosted interval. In momentum space this translates into the choice of the energy direction. In view of our pseudo-entropy interpretation we would then like to study the behaviour of our function (\ref{ententropy}) under rotations.

We saw that our two point function is zero unless the operators lie on a line centered at the origin, so the question is how the computation transforms under a rotation $k_1\rightarrow e^{i\theta}k_1$ of this line around the origin. One can easily see that
\begin{align}
 S(k_1,\lambda k_1)&=     -c\frac{8}{3} \;\pi \;\lim_{\epsilon^{R/I}\to 0}\delta(k_1^R\epsilon^I-k_1^I\epsilon^R)\,  |k_1|^2 F_{\pm}(\lambda,0) \notag \\
 &\rightarrow -c\frac{8}{3} \;\pi \;\lim_{\epsilon^{R/I}\to 0}\delta(\cos\theta\, [k_1^R\epsilon^I-k_1^I\epsilon^R]-\sin\theta\, [k_1^I\epsilon^I+k_1^R\epsilon^R])\,  | k_1|^2 F_{\pm}(\lambda,0)\,.
\end{align}

It will be convenient here to rewrite the delta function as the following limit
\begin{equation}
    \delta(x)=\lim_{\chi\to 0}\,\frac{1}{2}\,\chi\, |x|^{\chi-1}=\lim_{\chi\to 0}\, \frac{1}{2}\,\chi\,\frac{1}{|x|}+O(\chi^2)\,.
\end{equation}
Then temporary using the simple notation $k_1^R\epsilon^I-k_1^I\epsilon^R\equiv w$ and $k_1^I\epsilon^I+k_1^R\epsilon^R\equiv v$ we can assume a small value of the angle $\theta$ so that $|\sin\theta|\ll 1$ and expand the delta as:
\begin{align}
 \delta(\cos\theta\, [k_1^R\epsilon^I-k_1^I\epsilon^R]-\sin\theta\, [k_1^I\epsilon^I+k_1^R\epsilon^R])&=  \lim_{\chi\to 0}\,\frac{1}{2} \,\chi\left(\frac{1}{|\cos\theta\, w|}+\sum_{n=1}^{\infty}\frac{1}{|\cos\theta\, w|}\left(\frac{\sin\theta\, v}{\cos\theta\, w}\right)^n\right)+O(\chi^2)\notag \\
 &=\frac{\delta(k_1^R\epsilon^I-k_1^I\epsilon^R)}{|\cos\theta|}\left(1-\tan\theta\,\frac{k_1^I\epsilon^I+k_1^R\epsilon^R}{k_1^R\epsilon^I-k_1^I\epsilon^R}\right)^{-1}\,,
\end{align}
where in the last passage we have assumed $|\tan\theta \frac{v}{w}|<1$. Under these assumptions, it follows that
\begin{align}
 S(k_1,\lambda k_1)&\rightarrow S(e^{i\theta}k_1,\lambda e^{i\theta} k_1) =\frac{1}{|\cos\theta|}\left(1-\tan\theta\,\frac{k_1^I\epsilon^I+k_1^R\epsilon^R}{k_1^R\epsilon^I-k_1^I\epsilon^R}\right)^{-1}S(k_1,\lambda k_1) \,.
\end{align}

Let us now consider the simplest case of $k_1\in\mathbb{R}$. Then we obtain the simpler form
\begin{equation}
 S(k_1,\lambda k_1)_{k_1\in\mathbb{R}}=  -c\frac{8}{3} \;\pi \;\lim_{\epsilon^I\to 0}\delta(k_1^R\epsilon^I)\,  |k_1|^2 F_{\pm}(\lambda,0)=-c\frac{8}{3} \;\pi \;\lim_{\epsilon^I\to 0}\delta(\epsilon^I)\, |k_1| F_{\pm}(\lambda,0)\,.
\end{equation}
 Under a small boost, $k_1$ acquires a small imaginary part $k_1\,\sin\theta\ll 1$ so that
 \begin{equation}
 S(e^{i\theta}k_1,\lambda e^{i\theta} k_1)_{k_1\in\mathbb{R}}=-c\frac{8}{3} \;\pi \;\lim_{\epsilon^I\to 0}\delta(\epsilon^I)\,  |k_1| F_{\pm}(\lambda,0)\frac{1}{|\cos\theta|}\left(1-\tan\theta\,\frac{\epsilon^R}{\epsilon^I}\right)^{-1}\,.
\end{equation}
Note that picking $\lambda=-1$ the choice $\theta=\pi$ corresponds to the case of exchange of the two interval points analized in the previous subsection, under which we have invariance, which is consistent with the above result having $\tan\pi=0$.

Note that the function (\ref{ententropy}) is \emph{not} invariant under rigid rotations because of the regularization chosen inside the delta functions which breaks conformal invariance in momentum space, and in this case rotational invariance. We can choose a regularization that transforms under rotations so to keep (\ref{ententropy}) invariant by observing that, as the delta function argument can be rewritten as $k_1^R\epsilon^I-k_1^I\epsilon^R=\text{Im}(\bar{k}_1 \epsilon)$, sending $\epsilon\rightarrow e^{i\theta}\epsilon$ indeed does the job.

 \section{Conclusions} \label{sec:conclusions}
 We have proposed a definition for the twist operator acting on momentum space coordinates starting with an ansatz for its Fourier transform acting on a probe operator and then working out its formal properties. This classical derivation has been then upgraded to a quantum relation thus defining the two-point function of twist operators in momentum space as a certain integral over momenta of the standard Fourier transform of the position-space two-point function. This object has been computed explicitly and some of its properties studied in detail. Moreover we speculate on a possible interpretation of the result as a particular case of the pseudo-R\'enyi entropies recently introduced in \cite{Nakata:2020fjg}.
 
 Many questions remain open, most of them on the precise physical meaning of the object we proposed. In particular if a pseudo-R\'enyi entropy interpretation makes any sense at all, the derivation of the states involved in the reduced density matrix should be possible. Furthermore an understanding of the unusual constraints on the general form of the twist operator two point function has to be further studied. Finally, the connection with existing results for momentum space entanglement in the literature is yet to be understood in detail.

 \section*{Acknowledgments}
We thank Dmitry Melnikov and M\'at\'e Lencs\'es % and Diego Trancanelli
for comments and discussions. We also acknowledge financial support from MEC and MCTIC.

\appendix
\numberwithin{equation}{section}

\section{Integration in polar coordinates for $\lambda>0$}\label{poslsec}

Here we calculate the integral $I=I(k_1,k_2;\Delta)$ appearing in (\ref{eq:Renyi}). We do this by going to polar coordinates $k^R=r\cos\theta,k^I=r\sin\theta$ and $k'^R=r'\cos\theta',k'^I=r'\sin\theta'$ such that the integral becomes
\begin{align}\label{eq:Ipolar}
I = \int drd\theta dr'd\theta'\,\frac{1}{rr'}\,\delta(p_1+p_2)\,|p_1|^{2(\Delta+1)}\,,
\end{align}
where 
\begin{align}\label{eq:p1p2polar}
p_1 &= 4\left(r\cos\theta-r^2\right)k_1\qquad\qquad 
p_2 = 4\left(r'\cos\theta'-r'^2\right)k_2\,.
\end{align}

Now the way in which we proceed with the integration of the delta functions will depend on the form of the external momenta $k_1,k_2$. We first need to recall that $\delta(p_1+p_2)=\delta(p_1^R+p_2^R)\delta(p_1^I+p_2^I)$ and the well-known property $\delta[f(x)]=\sum_{x_*}\frac{\delta(x-x_*)}{|\partial_xf(x_*)|}$ with $x_*$ such that $f(x_*)=0$. 
It is crucial to note that, since the prefactors multiplying $k_1$ and $k_2$ in \eq{p1p2polar} are both real-valued, we have only two possibilities for reaching $p_1+p_2=0$ in the delta argument:
\begin{enumerate}
    \item{$k_1,k_2$ colinear, i.e., $k_2=\lambda\,k_1$ with $\lambda\in\mathbb{R}$ ($\lambda\ne1$ to avoid insertions at the same point):\\
    In other words, $k_2^R/k_1^R = k_2^I/k_1^I = \lambda$. 
    In this case the real and imaginary parts of the delta function arguments become proportional: $\delta(p_1+p_2)=\delta\big(4 k_1^R f_{\lambda}(r,r',\theta,\theta')\big)\delta\big((k_1^I/k_1^R)4 k_1^R f_{\lambda}(r,r',\theta,\theta')\big)$. To find the delta zeros we have then to solve the quadratic equation
\begin{align}\label{quadeqi}
f_{\lambda}(r,r',\theta,\theta')=\left(r\cos\theta-r^2\right) + \lambda\left(r'\cos\theta'-r'^2\right)= 0\,.
\end{align} 
for either one of the integration variables $r,\theta,r',\theta'$. 

}

\item{$k_1,k_2$ not colinear:\\
In this case the only possibility of $p_1+p_2=0$ is to set $p_1$ and $p_2$ to zero separately.
Because of the factor of $|p_1|^{2(\Delta+1)}$ this will give a vanishing integral for any $\Delta>-1$, meaning in particular that the interesting region $n\geq 1$ is excluded \footnote{We are here looking at the usual $n=1$ limit in analogy with position-space where it is used to extract the entanglement entropy. More on this will be discussed later on. }. Because of this we will concentrate on the collinear case.
}
\end{enumerate}
Since the two delta functions are proportional, we necessarily end up with a divergent result that should be regularized. In order to do so we rewrite the product of deltas as $\delta(p_1^R+p_2^R)\delta(p_1^I+p_2^I)\rightarrow \delta(p_1^R+p_2^R-\epsilon^R)\delta(p_1^I+p_2^I-\epsilon^I)$. The delta function can then be written either as
\begin{equation}
  \delta(p_1+p_2-\epsilon) = \lim_{\epsilon^{R/I}\to 0}\delta(4k_1^R f_{\lambda}-\epsilon^R) \delta(\frac{k_1^I}{k_1^R}\epsilon^R-\epsilon^I)=\frac{ \delta(x-x_*)}{4|k_1^R f_{\lambda}'(x_*)|}\lim_{\epsilon^{R/I}\to 0}\delta(\frac{k_1^I}{k_1^R}\epsilon^R-\epsilon^I)
\end{equation}
or as
\begin{equation}
  \delta(p_1+p_2-\epsilon) = \lim_{\epsilon^{R/I}\to 0}\delta(4k_1^I f_{\lambda}-\epsilon^I) \delta(\frac{k_1^R}{k_1^I}\epsilon^I-\epsilon^R)=\frac{ \delta(x-x_*)}{4|k_1^I f_{\lambda}'(x_*)|}\lim_{\epsilon^{R/I}\to 0}\delta(\frac{k_1^R}{k_1^I}\epsilon^I-\epsilon^R)\; ,
\end{equation}
where the derivative $f'$ is with respect to whatever variable $x$ we are using to solve the delta argument. % and evaluated at the solution(s) $x=x_*$  $f_{\lambda}(x_*)=0$,
The two results agree by extracting a constant from the regulated delta, obtaining
\begin{equation}
   \delta(p_1+p_2-\epsilon) = \frac{ \delta(x-x_*)}{4|f'_{\lambda}(x_*)|}\lim_{\epsilon^{R/I}\to 0}\delta(k_1^R\epsilon^I-k_1^I\epsilon^R)\; .
\end{equation}
In the following we will loosely refer to the regulated delta as $\delta(k_1^R\epsilon^I-k_1^I\epsilon^R)\sim \delta(0)$.%, unless where needed the more refined notation.

For $\lambda=0$ we can solve (\ref{quadeqi}) in the variable $x=r$, obtaining the solutions
\begin{align}\label{solr}
r_{*,\pm}=  \frac{1}{2}\left(\cos\theta \pm \sqrt{\cos^2\theta+4 r' \lambda (\cos\theta'-r')}\right)
\equiv \frac{1}{2}\left(\cos\theta \pm T(r',\theta',\theta)\right)\,.
\end{align} 
For $\lambda\neq 0$ instead we solve (\ref{quadeqi}) in $x=r'$ to obtain
\begin{align}\label{solrp}
r'_{*,\pm}=  \frac{1}{2}\left(\cos\theta' \pm \sqrt{\cos^2\theta'+4\frac{r}{\lambda} (\cos\theta-r)}\right)
\equiv \frac{1}{2}\left(\cos\theta' \pm R(r,\theta,\theta')\right)\,.
\end{align} 
We concentrate now on (\ref{solrp}) as the $\lambda\rightarrow1/\lambda$ transformation is part of the $k_1\rightarrow k_2$ symmetry to be discussed.
The domain of $r$ is restricted by the requirement
\begin{equation}
    \cos^2\theta'+4\lambda^{-1}r(\cos\theta-r)\geq 0\;,
\end{equation}
which is solved by:
\begin{itemize}
   \item $\lambda>0 \;\;\;\;\; 0\leq r\leq \frac{1}{2}\left(\cos\theta+\sqrt{\cos^2\theta +\lambda \cos^2\theta'}\right) $
   \item $ \lambda<0, \; \cos\theta>0 \;\;\;\;\; 0\leq r\leq \frac{1}{2}\left(\cos\theta-\sqrt{\cos^2\theta +\lambda \cos^2\theta'}\right) \;\; \text{and} \;\;  \frac{1}{2}\left(\cos\theta+\sqrt{\cos^2\theta +\lambda \cos^2\theta'}\right)\leq r$
    \item  $\lambda<0, \; \cos\theta<0 \;\;\;\;\; 0\leq r$ ,
\end{itemize}
with the additional condition for negative $\lambda$ that $\lambda>-\frac{\cos^2\theta}{\cos^2\theta'}$. 
We can then see that the existence of \emph{positive} roots $r'_{*,\pm}$  depends on the value of $r$ as well as the signs of $\lambda$, $\cos\theta$ and $\cos\theta'$. For reasons to be explained later we fix $\lambda>0$ and treat the negative case in the appendix. Then the delta function argument has the following roots:

   \begin{itemize} 
   \item $\cos\theta'\geq 0,\;\cos\theta>0 \;\;\;\;\;\; r'_{*,+} \;  \text{for} \; 0 \leq r < \cos\theta \;\;\;\;
    r'_{*,\pm} \;  \text{for} \; \cos\theta \leq r \leq  \frac{1}{2}\left(\cos\theta+\sqrt{\cos^2\theta +\lambda \cos^2\theta'}\right) $
     \item $\cos\theta'\geq 0,\;\cos\theta<0 \;\;\;\;\;\; r'_{*,\pm} \;\text{for}\; 0\leq r\leq \frac{1}{2}\left(\cos\theta+\sqrt{\cos^2\theta +\lambda \cos^2\theta'}\right)   $
 \item $\cos\theta'\leq 0,\; \cos\theta>0 \;\;\;\;\;\; r'_{*,+} \;  \text{for} \; 0 \leq r < \cos\theta $
    \item $\cos\theta'\leq 0,\; \cos\theta<0\;\;\;\;\;\;\text{no} \; \text{roots}\,.$
            \end{itemize}

The integral \eq{Ipolar}, which so far reads
\begin{align}\label{integ}
I &= \frac{\delta(0)\,(4|k_1|)^{2(\Delta+1)}}{4|\lambda|}\int drd\theta dr'd\theta'\,\frac{(r\cos\theta-r^2)^{2(\Delta+1)}}{rR(r,\theta,\theta')} \frac{1}{r'}\sum_{i=\pm} \delta(r'-r'_{*,i})\,,
\end{align}
then acquires a complicated structure in its integration domain by following the above formulas. Abbreviating the integrand as $f(r')$, we have
\begin{align}\label{uno}
   I &\propto \int_{-\frac{\pi}{2}}^{\frac{\pi}{2}}d\theta'\; \int_{-\frac{\pi}{2}}^{\frac{\pi}{2}}d\theta\; \left(\int_{0}^{\cos\theta} \hspace{-0.3cm}dr f(r'_{*,+})+\int_{\cos\theta}^{\frac{1}{2}(\cos\theta+\sqrt{\cos^2\theta +\lambda \cos^2\theta'})}\hspace{-3cm}dr (f(r'_{*,+})+f(r'_{*,-}))\right)\notag \\
    & +\int_{-\frac{\pi}{2}}^{\frac{\pi}{2}}d\theta'\; \int_{\frac{\pi}{2}}^{\frac{3\pi}{2}}d\theta\; \int_{0}^{\frac{1}{2}(\cos\theta+\sqrt{\cos^2\theta +\lambda \cos^2\theta'})} \hspace{-3cm}dr (f(r'_{*,+})+f(r'_{*,-}))  +\int_{\frac{\pi}{2}}^{\frac{3\pi}{2}}d\theta'\; \int_{-\frac{\pi}{2}}^{\frac{\pi}{2}}d\theta\; \int_{0}^{\cos\theta} \hspace{-0.3cm}dr f(r'_{*,+}) \;,
\end{align}
where the integrands differ depending if we have a single ($r'_{*,+}$) or two ($r'_{*,\pm}$) roots for the delta function argument, that is
\begin{align}\label{integ1}
\int drd\theta d\theta'\,f(r'_{*,+})=2\int drd\theta d\theta'\,\frac{(r^2-r\cos\theta)^{2(\Delta+1)}}{rR(r,\theta,\theta')\left(\cos\theta' + R(r,\theta,\theta')\right)} 
\end{align}
and
\begin{align}\label{integ2}
\int drd\theta d\theta'\,(f(r'_{*,+})+f(r'_{*,-}))=\lambda\int drd\theta d\theta'\,\frac{(r^2-r\cos\theta)^{2\Delta+1}\cos\theta'}{rR(r,\theta,\theta')} \;.
\end{align}
As a function of $\theta'$ the integrand of (\ref{integ1}) is an odd function of $\cos\theta'$ plus a $\theta'-$independent part, %($=\frac{\lambda}{4}(r\cos\theta-r^2)^{-1}$)
\[
\frac{1}{R(r,\theta,\theta')(\cos\theta' + R(r,\theta,\theta'))}=\frac{\lambda}{4r(\cos\theta-r)}+ f_{\text{odd}}( \cos\theta')
\]
while for (\ref{integ2}) it is just an odd function
\[
\frac{\cos\theta'}{R(r,\theta,\theta')}= g_{\text{odd}}( \cos\theta')\; .
\]
This leads to some simplifications in the first and second line of (\ref{uno}):
\begin{align}\label{sim1}
   & \left(\int_{-\frac{\pi}{2}}^{\frac{\pi}{2}}d\theta'+\int_{\frac{\pi}{2}}^{\frac{3\pi}{2}}d\theta'\right) \int_{-\frac{\pi}{2}}^{\frac{\pi}{2}}d\theta\; \int_{0}^{\cos\theta} \hspace{-0.3cm}dr f(r'_{*,+})= \lambda\pi\int_{-\frac{\pi}{2}}^{\frac{\pi}{2}}d\theta\int_{0}^{\cos\theta}  \hspace{-0.3cm}dr  \frac{(r\cos\theta-r^2)^{2\Delta+1}}{r}  \,.
\end{align}
We then need to compute
\begin{align}\label{unosim}
 I=  &  \frac{\delta(0)\,(4|k_1|)^{2(\Delta+1)}}{4}\left[ \int_{-\frac{\pi}{2}}^{\frac{\pi}{2}}\hspace{-0.2cm}d\theta'\left(\int_{-\frac{\pi}{2}}^{\frac{\pi}{2}}\hspace{-0.2cm}d\theta \int_{\cos\theta}^{f_+}\hspace{-0.2cm}dr \frac{(r^2-r\cos\theta)^{2\Delta+1}\cos\theta'}{r\sqrt{\cos^2\theta'+4\frac{r}{\lambda} (\cos\theta-r)}}\right.\right.\notag \\
    & \left.\left.+ \int_{\frac{\pi}{2}}^{\frac{3\pi}{2}}\hspace{-0.2cm}d\theta \int_{0}^{f_+} \hspace{-0.2cm}dr \frac{(r^2-r\cos\theta)^{2\Delta+1}\cos\theta'}{r\sqrt{\cos^2\theta'+4\frac{r}{\lambda} (\cos\theta-r)}}\right) + \pi\int_{-\frac{\pi}{2}}^{\frac{\pi}{2}}d\theta\int_{0}^{\cos\theta}  \hspace{-0.3cm}dr  \frac{(r\cos\theta-r^2)^{2\Delta+1}}{r}\right] \;,
\end{align}
having defined
\[
f_+(\cos\theta,\cos\theta')=\frac{1}{2}(\cos\theta+\sqrt{\cos^2\theta +\lambda \cos^2\theta'}) 
\]
and
\[
f_-(\cos\theta,\cos\theta')=\frac{1}{2}(\cos\theta-\sqrt{\cos^2\theta +\lambda \cos^2\theta'}) \;.
\]

The integrals in (\ref{unosim}) can be solved at $\Delta=0$.\footnote{As we have already partially discussed our two point function is not, in general, associated with a Renyi entropy in momentum space. Because of this we cannot extract a corresponding entanglement entropy by partial derivative in $n$ at $n=1$. Nevertheless the analogy with Renyi entropies may remain useful, as we will discuss, and in the present context the interesting (and computable) limit is $\Delta=0$. } We have two types of terms: a simple double integral 
\begin{equation}\label{di}
\pi\int_{-\frac{\pi}{2}}^{\frac{\pi}{2}}d\theta\int_{0}^{\cos\theta}  \hspace{-0.3cm}dr  (\cos\theta-r)=\frac{\pi^2}{4}\,,
\end{equation}
and a slightly more complicated triple integral that, by a simple coordinate transformation, can be brought to the form
\[
\int_{-\frac{\pi}{2}}^{\frac{\pi}{2}}\hspace{-0.2cm}d\theta'\int_{-\frac{\pi}{2}}^{\frac{\pi}{2}}\hspace{-0.2cm}d\theta \left(\int_{\cos\theta}^{f_+}\hspace{-0.2cm}dr \frac{(r-\cos\theta)\cos\theta'}{\sqrt{\cos^2\theta'+4\frac{r}{\lambda} (\cos\theta-r)}}+ \int_{0}^{f_+(-\cos\theta)} \hspace{-0.6cm}dr \frac{(r+\cos\theta)\cos\theta'}{\sqrt{\cos^2\theta'-4\frac{r}{\lambda} (\cos\theta+r)}}\right)=\frac{1}{4}\pi^2\lambda \,.
\]
Adding (\ref{di}) we finally obtain
\begin{equation}\label{poslamb}
 I(\Delta=0)= \delta(0)\,|k_1|^2\pi^2(\lambda+1)\,.
\end{equation}

\section{Integration in spherical coordinates for $\lambda>0$}\label{possphsec}

\begin{figure}[t]
\centering
\includegraphics[width=.7\textwidth]{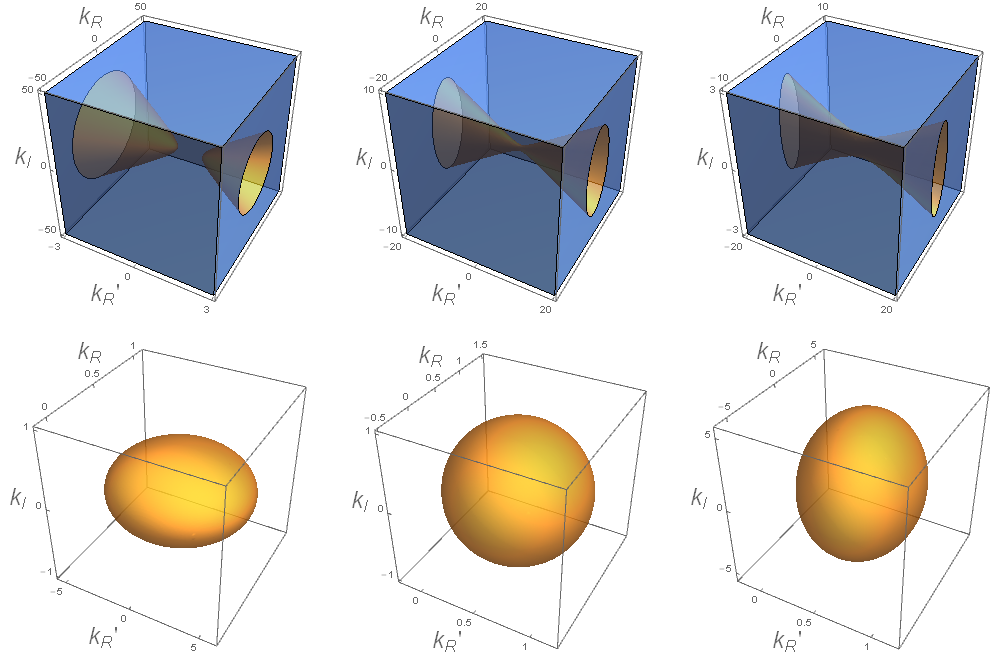}
\caption{Integration region $\mathcal{R}(\lambda)$ for different values of $\lambda$. Top: $\lambda=-100,-1,-0.01$, respectively. Bottom: $\lambda=0.01,2,100$, respectively. 
For negative $\lambda$ the region is $\mathbb{R}^3$ with a solid hyperboloid removed (for $\lambda<-1$ the hyperboloid becomes two-sheeted). For positive $\lambda$ the region is the interior of an ellipsoid which is oblate for $0<\lambda<1$ and prolate for $\lambda>1$.}
\label{fig:regions}
\end{figure}

Here we provide an alternative calculation of the integral $I=I(k_1,k_2;\Delta)$ appearing in (\ref{eq:Renyi}) using a different coordinate change. Using cartesian coordinates for each integration momenta, the integral reads
\begin{align}\label{eq:Icartesian}
I &= (4|k_1|)^{2(\Delta+1)}\int dk^Rdk^Idk'^Rdk'^I\,\frac{\big[k^R-(k^R)^2-(k^I)^2\big]^{2(\Delta+1)}}{\big[(k^R)^2+(k^I)^2\big]\big[(k'^R)^2+(k'^I)^2\big]}\,\delta(p_1+p_2)\,,
\end{align}
where $\delta(p_1+p_2)=\delta(p_1^R+p_2^R)\,\delta(p_1^I+p_2^I)$ and 
\begin{align}\label{eq:p1p2cartesian}
p_1+p_2 = 4\left(k^R-(k^R)^2-(k^I)^2\right)k_1+4\left(k'^R-(k'^R)^2-(k'^I)^2\right)k_2\,.
\end{align}
Again the first thing to do is integrate the delta function; as in polar coordinates the only non-vanishing case is for  $k_2=\lambda\,k_1$ ($\lambda\in\mathbb{R}$). Solving the delta for $k'^I$ we get
\begin{align}\label{eq:deltaroots}
k'^I_{*,\pm} &= \pm\sqrt{k'^R-(k'^R)^2 +\lambda^{-1}\left(k^R-(k^R)^2-(k^I)^2\right)} \equiv \pm\sqrt{g_\lambda(k^R,k^I,k'^R)}
\end{align}  
and therefore the integral \eq{Icartesian} then becomes
\begin{align}
I &= \frac{\delta(0)}{4}(4|k_1|)^{2(\Delta+1)}
\int dk^Rdk^Idk'^Rdk'^I\,\frac{\big[k^R-(k^R)^2-(k^I)^2\big]^{2(\Delta+1)}}{\big[(k^R)^2+(k^I)^2\big]\big[(k'^R)^2+(k'^I)^2\big]}\sum_{i=\pm}\frac{\delta(k'^I-k'^I_{*,i})}{2|k'^I_{*,i}|}
\end{align}
Here we should pause for a moment to discuss the integration region. In principle, each integration variable runs over the full real line. However, integration of the delta function on $k'^I$ only yields a non-vanishing result for values of $k^R,k^I,k'^R$ such that the delta roots \eq{deltaroots} are real-valued, namely the region
\begin{align}\label{eq:constraint}
\mathcal{R}(\lambda)\equiv\left\{\big(k^R,k^I,k'^R\big)\in\mathbb{R}^3\,\big|\,g_\lambda(k^R,k^I,k'^R)\ge0\right\}\,.
\end{align}
In order to identify this region, we first rewrite $f_\lambda(k^R,k^I,k'^R)$ as
\begin{align}
g_\lambda(k^R,k^I,k'^R) = -\lambda^{-1}\left[\lambda\left(k'^R-\frac{1}{2}\right)^2+\left(k^R-\frac{1}{2}\right)^2+\left(k^I\right)^2-\frac{1+\lambda}{4}\right]\,.
\end{align}
Then it becomes clear that for positive $\lambda$ the region $\mathcal{R}(\lambda)$ is the interior of an ellipsoid (oblate for $0<\lambda<1$ and prolate for $\lambda>1$). For negative $\lambda$, $\mathcal{R}(\lambda)$ is the exterior of a hyperboloid (one-sheeted for $-1\le\lambda<0$ and two-sheeted for $\lambda<-1$) in 3-dimensional space. This is illustrated in figure \fig{regions}.
Then
\begin{align}
I &= \frac{\delta(0)}{4}(4|k_1|)^{2(\Delta+1)}\frac{1}{|\lambda|}
\int_{\mathcal{R}(\lambda)} dk^Rdk^Idk'^R\,\frac{\big[k^R-(k^R)^2-(k^I)^2\big]^{2(\Delta+1)}}{\big[(k^R)^2+(k^I)^2\big]\big[(k'^R)^2+f_\lambda(k^R,k^I,k'^R)\big]\sqrt{g_\lambda(k^R,k^I,k'^R)}}\,
\end{align}
and we recall that $g_\lambda(k^R,k^I,k'^R)$ is defined in \eq{deltaroots}.

For $\lambda>0$, we redefine the coordinates ($k^R,k^I,k'^R$) as  
\begin{align}
k^R = \frac{1}{2}+\frac{x}{2}\,,\qquad\qquad 
k^I = \frac{y}{2}\,,\qquad\qquad 
k'^R = \frac{1}{2}+\frac{z}{2\sqrt{\lambda}}
\end{align}
so that the integration region becomes a ball of radius $R_\lambda\equiv\sqrt{1+\lambda}$, which we denote by $\mathcal{B}(R_\lambda)$. In this case $g_\lambda(k^R,k^I,k'^R)$ becomes $g_\lambda(x,y,z) = -\frac{1}{4\lambda}\left(x^2+y^2+z^2-(1+\lambda)\right)$ and the integral to solve is
\begin{align}
I &= 
 \frac{\delta(0)}{4}(4|k_1|)^{2(\Delta+1)}\int_{\mathcal{B}(R_\lambda)} dxdydz\,\frac{4^{-2\Delta-1}\lambda^{-1}\big[2(1+x)-(1+x)^2-y^2\big]^{2(\Delta+1)}}{\big[(1+x)^2+y^2\big]\big[(1+\frac{z}{\sqrt{\lambda}})^2+\lambda^{-1}\left(1+\lambda-x^2-y^2-z^2\right)\big]\sqrt{1+\lambda-x^2-y^2-z^2}}\,.
\end{align}
Of course the natural way to integrate over a spherical region is to use spherical coordinates,
\begin{align}
I 
&=  \frac{\delta(0)}{4}(4|k_1|)^{2(\Delta+1)}\int_{0}^{R_\lambda} dr \int_{0}^{\pi}d\theta \int_{0}^{2\pi}d\phi\,\frac{4^{-2\Delta-1}\lambda^{-1}r^2\sin\theta \big[1-r^2\sin^2\theta\big]^{2(\Delta+1)}}{\big[1+2r\sin\theta\cos\phi+r^2\sin^2\theta\big]\big[(1+\frac{r\cos\theta}{\sqrt{\lambda}})^2+\lambda^{-1}\left(R_\lambda^2-r^2\right)\big]\sqrt{R_\lambda^2-r^2}}\,.
\end{align}
The $\phi$ integral can be done using the identity %(see Gradshtein\,\&\,Ryzhik)
\begin{align}\label{eq:fi_integral}
\int_0^{2\pi} \frac{d\phi}{1-2a\cos\phi+a^2}&=
\begin{cases}
\frac{2\pi}{1-a^2}, & a^2<1\\
\frac{2\pi}{a^2-1}, & a^2>1
\end{cases}
=\text{sgn}(a^2-1)\frac{2\pi}{a^2-1}
\end{align}
with $a=-r\sin\theta$. In our case we have to be careful and split integrals since $a^2$ can fall either below or above $1$ depending on the ranges of integration. Then we have
\begin{align}
I &= 
-2\pi  \frac{\delta(0)}{4}(4|k_1|)^{2(\Delta+1)} \frac{4^{-2\Delta-1}}{\lambda}\int_{0}^{R_\lambda} dr \int_{0}^{\pi}d\theta \,\,\text{sgn}(r^2\sin^2\theta-1)\frac{r^2\sin\theta \big[1-r^2\sin^2\theta\big]^{2\Delta+1}}{\big[(1+\frac{r\cos\theta}{\sqrt{\lambda}})^2+\lambda^{-1}\left(R_\lambda^2-r^2\right)\big]\sqrt{R_\lambda^2-r^2}}\,.
\end{align}
% \begin{align}
% I_\Delta(\lambda>0) &= 
% -2\pi \frac{4^{-2\Delta-1}}{\lambda}\int_{0}^{R_\lambda} dr \int_{0}^{\pi}d\theta \,\frac{r^2\sin\theta \big[1-r^2\sin^2\theta\big]^{2\Delta+1}}{\big[(1+\frac{r\cos\theta}{\sqrt{\lambda}})^2+\lambda^{-1}\left(R_\lambda^2-r^2\right)\big]\sqrt{R_\lambda^2-r^2}}\,.
% \end{align}
The $\theta$ integral gets nicer by changing variables to $u=\cos\theta$,
\begin{align}
I &= 
-2\pi  \frac{\delta(0)}{4}(4|k_1|)^{2(\Delta+1)}\frac{4^{-2\Delta-1}}{\lambda}\int_{0}^{R_\lambda} dr \int_{-1}^{1}du \,\,\text{sgn}(r^2(1-u^2)-1)\frac{r^2\big[1-r^2+r^2u^2\big]^{2\Delta+1}}{\big[(1+\frac{ru}{\sqrt{\lambda}})^2+\lambda^{-1}\left(R_\lambda^2-r^2\right)\big]\sqrt{R_\lambda^2-r^2}}
\,.
\end{align}
%\footnote{Note that $\text{sgn}(r^2(1-u^2)-1)=+1$ for $r>1,u\in(-\frac{\sqrt{r^2-1}}{r},\frac{\sqrt{r^2-1}}{r})$ or $=-1$ otherwise.}% for $\{r<1,u\in(-1,1)\}\text{ or }\{r>1,u<-\frac{\sqrt{r^2-1}}{r}\}\text{ or }\{r>1,u>\frac{\sqrt{r^2-1}}{r}\}$.)
In order to proceed, we first note that
\begin{align*}
\text{sgn}[r^2(1-u^2)-1] = 
\begin{cases}
+1\qquad \text{if}\quad (r>1\text{ and }|u|<\sqrt{1-r^{-2}})\\
-1\qquad \text{if}\quad (r<1)\quad\text{or}\quad (r>1\text{ and }|u|>\sqrt{1-r^{-2}})
\end{cases}
\end{align*}
so the integrals can be split as 
\begin{align}
I
&= 
-2\pi  \frac{\delta(0)}{4}(4|k_1|)^{2(\Delta+1)}\frac{4^{-2\Delta-1}}{\lambda}\left[\int_{1}^{R_\lambda} dr \left(-\int_{-1}^{-\sqrt{1-r^{-2}}}du\,h_\Delta(r,u)+\int_{-\sqrt{1-r^{-2}}}^{\sqrt{1-r^{-2}}}du\,h_\Delta(r,u)-\int_{\sqrt{1-r^{-2}}}^{1}du\,h_\Delta(r,u)\right)\right.\notag\\
&\left.\qquad\qquad\qquad-\int_{0}^{1} dr \int_{-1}^{1}du\,h_\Delta(r,u)\right]\notag\\
&= 
-2\pi  \frac{\delta(0)}{4}(4|k_1|)^{2(\Delta+1)}\frac{4^{-2\Delta-1}}{\lambda}\left[-\int_{0}^{R_\lambda} dr \int_{-1}^{1}du\,h_\Delta(r,u)+2\int_{1}^{R_\lambda}\int_{-\sqrt{1-r^{-2}}}^{\sqrt{1-r^{-2}}}du\,h_\Delta(r,u)\right]
\,,
\end{align}
where for simplicity we introduced
\begin{equation}
h_\Delta(r,u)  \equiv\frac{r^2\big[1-r^2+r^2u^2\big]^{2\Delta+1}}{\big[(1+\frac{ru}{\sqrt{\lambda}})^2+\lambda^{-1}\left(R_\lambda^2-r^2\right)\big]\sqrt{R_\lambda^2-r^2}}\,.
\end{equation}
The $u$ integral above is still hard for arbitrary $\Delta$, but fortunately our case of interest $\Delta=0$ it can be done. The result after some simplifications is the following
\begin{align}
I(\Delta=0) &= 
-\frac{2\pi}{\lambda} \delta(0)|k_1|)^2\left[-\int_{0}^{R_\lambda} dr \,\frac{\lambda\,r}{\sqrt{R_\lambda^2-r^2}}\left(2r+\sqrt\lambda\log\frac{1+2\lambda-2\sqrt\lambda r}{1+2\lambda+2\sqrt\lambda r}\right) \right.\notag\\
&\left.\qquad\qquad +2\int_{1}^{R_\lambda} dr \,\frac{\lambda\,r}{\sqrt{R_\lambda^2-r^2}}\left(2\sqrt{r^2-1}
+\sqrt\lambda\log\frac{\lambda-\sqrt{\lambda(r^2-1)}}{\lambda+\sqrt{\lambda(r^2-1)}}\right)\right]
\,.
\end{align}
%Then we do $r\to r'=\frac{r}{R_\lambda}$ and finally solve it:
which finally can be integrated to give
\begin{align}\label{ilo}
I(\Delta=0)  
% &= 
% -\frac{2\pi}{4\lambda}\sqrt{1+\lambda}\left[-\int_{0}^{1} dr' \,\frac{2\lambda\,r'}{\sqrt{1-r'^2}}\left(\sqrt{1+\lambda}\,r'-\sqrt{\lambda}\,\text{arctanh}\left(\frac{2\sqrt{\lambda(1+\lambda)}}{1+2\lambda}r'\right)\right)\right.\notag\\
% &\left.\qquad\qquad\qquad +4\int_{1/\sqrt{1+\lambda}}^{1} dr' \,\frac{\lambda\,r'}{\sqrt{1-r'^2}}\left(\sqrt{(1+\lambda)r'^2-1}
% -\sqrt\lambda\,\text{arctanh}\left(\sqrt{r'^2+(r'^2-1)/\lambda}\right)\right)\right]\notag\\
&= 
-\frac{2\pi}{\lambda}\delta(0)|k_1|^2\left[\frac{\pi\lambda(\lambda-1)}{2}-\pi\lambda^2\right]=
\delta(0)|k_1|^2\pi^2(1+\lambda)
% &= 
% \frac{\pi^2}{4}(\lambda-1)
\,.
\end{align}

\section{$\lambda<0$}\label{polarnegl}

Let us now repeat the computation of section (\ref{poslsec}) for negative $\lambda$. The roots of (\ref{solrp}) are now
\begin{itemize}
   
         \item $\cos\theta'>0 \;\; \cos\theta>0$ 
         \begin{align}
      &r'_{*,\pm} \;  \text{for} \;   0\leq r\leq \frac{1}{2}\left(\cos\theta-\sqrt{\cos^2\theta +\lambda \cos^2\theta'}\right) \; \text{and}\;\frac{1}{2}\left(\cos\theta+\sqrt{\cos^2\theta +\lambda \cos^2\theta'}\right)\leq r\leq \cos\theta  \notag \\
       &r'_{*,+} \;  \text{for} \;  \cos\theta \leq r
       \end{align}
            \item $\cos\theta'>0 \;\; \cos\theta<0 \;\;\;\;\; r'_{*,+} \; \text{for} \; 0\leq r  $
         \item $\cos\theta'<0 \;\; \cos\theta>0\;\;\;\;\;   r'_{*,+} \;  \text{for} \; \cos\theta\leq r$
            \item $\cos\theta<0 \;\; \cos\theta<0\;\;\;\;\; r'_{*,+} \;\text{for}\; 0\leq r  $
\end{itemize}

Compared to the positive $\lambda$ case the integral \eq{Ipolar} has an additional complication
 due to the constrain $\cos^2\theta\geq -\lambda\cos^2\theta'$, thus restricting the integral domain of $\theta'$ :
\begin{align}\label{due}
   I&\propto \int_{-\frac{\pi}{2}}^{\frac{\pi}{2}}d\theta\;\left(\int_{-\frac{\pi}{2}}^{-\arccos\frac{\cos\theta}{\sqrt{-\lambda}}}\hspace{-1.0cm}d\theta'+ \int_{\arccos\frac{\cos\theta}{\sqrt{-\lambda}}}^{\frac{\pi}{2}}\hspace{-1.0cm}d\theta'\hspace{0.4cm}\right)\;  \left(\int_{0}^{\frac{1}{2}(\cos\theta-\sqrt{\cos^2\theta +\lambda \cos^2\theta'})} \hspace{-3cm}dr (f(r'_{*,+})+f(r'_{*,-})+\int_{\frac{1}{2}(\cos\theta+\sqrt{\cos^2\theta +\lambda \cos^2\theta'})}^{\cos\theta} \hspace{-3cm}dr (f(r'_{*,+})+f(r'_{*,-})+\int_{\cos\theta}^{\infty} \hspace{-0cm}dr f(r'_{*,+})\hspace{+0cm}\right)\notag \\
      & +\int_{\frac{\pi}{2}}^{\frac{3\pi}{2}}d\theta\;\left(\int_{-\frac{\pi}{2}}^{-\arccos\frac{-\cos\theta}{\sqrt{-\lambda}}}\hspace{-1.0cm}d\theta'+ \int_{\arccos\frac{-\cos\theta}{\sqrt{-\lambda}}}^{\frac{\pi}{2}}\hspace{-1.0cm}d\theta'\hspace{0.4cm}\right)\;  \int_{0}^{\infty} \hspace{-0cm}dr f(r'_{*,+}) + \int_{-\frac{\pi}{2}}^{\frac{\pi}{2}}d\theta\;\left(\int_{\frac{\pi}{2}}^{\arccos\frac{-\cos\theta}{\sqrt{-\lambda}}}\hspace{-1.0cm}d\theta'+\int_{\arccos\frac{\cos\theta}{\sqrt{-\lambda}}+\pi}^{\frac{3\pi}{2}}\hspace{-1.0cm}d\theta'\hspace{0.5cm}\right)\;\int_{\cos\theta}^{\infty} \hspace{-0cm}dr f(r'_{*,+}) \notag \\
   &+\int_{\frac{\pi}{2}}^{\frac{3\pi}{2}}d\theta\;\left(\int_{\frac{\pi}{2}}^{\arccos\frac{\cos\theta}{\sqrt{-\lambda}}}\hspace{-1.0cm}d\theta'+\int_{\arccos\frac{-\cos\theta}{\sqrt{-\lambda}}+\pi}^{\frac{3\pi}{2}}\hspace{-1.0cm}d\theta'\hspace{0.5cm}\right)\;  \int_{0}^{\infty} \hspace{-0.3cm}dr f(r'_{*,+}) \;,
\end{align}

Due to the symmetry of the integrand in $\theta'$ the four terms with integrand $f(r'_{*,+})$ in (\ref{due}) produce respectively
\begin{align}\label{sim2}
 \int_{-\frac{\pi}{2}}^{\frac{\pi}{2}}d\theta &\;\left(\int_{-\frac{\pi}{2}}^{-\arccos\frac{\cos\theta}{\sqrt{-\lambda}}}\hspace{-1.0cm}d\theta'+ \int_{\arccos\frac{\cos\theta}{\sqrt{-\lambda}}}^{\frac{\pi}{2}}\hspace{-1.0cm}d\theta'\hspace{0.4cm}+\int_{\frac{\pi}{2}}^{\arccos\frac{-\cos\theta}{\sqrt{-\lambda}}}\hspace{-1.0cm}d\theta'+\int_{\arccos\frac{\cos\theta}{\sqrt{-\lambda}}+\pi}^{\frac{3\pi}{2}}\hspace{-1.0cm}d\theta'\hspace{0.5cm}\right)\int_{\cos\theta}^{\infty} \hspace{-0cm}dr f(r'_{*,+}) \notag \\
 = &2\lambda\int_{-\frac{\pi}{2}}^{\frac{\pi}{2}}d\theta\, \left(\frac{\pi}{2}-\arccos\frac{\cos\theta}{\sqrt{-\lambda}}\right)\int_{\cos\theta}^{\infty} \hspace{-0cm}dr\,\frac{(r\cos\theta-r^2)^{2\Delta+1}}{r}
\end{align}
and 
\begin{align}\label{sim3}
 \int_{\frac{\pi}{2}}^{\frac{3\pi}{2}}d\theta &\;\left(\int_{-\frac{\pi}{2}}^{-\arccos\frac{-\cos\theta}{\sqrt{-\lambda}}}\hspace{-1.0cm}d\theta'+ \int_{\arccos\frac{-\cos\theta}{\sqrt{-\lambda}}}^{\frac{\pi}{2}}\hspace{-1.0cm}d\theta'\hspace{0.4cm}+\int_{\frac{\pi}{2}}^{\arccos\frac{\cos\theta}{\sqrt{-\lambda}}}\hspace{-1.0cm}d\theta'+\int_{\arccos\frac{-\cos\theta}{\sqrt{-\lambda}}+\pi}^{\frac{3\pi}{2}}\hspace{-1.0cm}d\theta'\hspace{0.5cm}\right)\int_{0}^{\infty} \hspace{-0cm}dr f(r'_{*,+}) \notag \\
 = &2\lambda\int_{\frac{\pi}{2}}^{\frac{3\pi}{2}}d\theta\, \left(\frac{\pi}{2}-\arccos\frac{-\cos\theta}{\sqrt{-\lambda}}\right)\int_{0}^{\infty} \hspace{-0cm}dr\,\frac{(r\cos\theta-r^2)^{2\Delta+1}}{r} \,.
\end{align}
The sum of the two gives
\begin{align}\label{simsum1}
& 2\lambda\int_{-\frac{\pi}{2}}^{\frac{\pi}{2}}d\theta\, \left(\frac{\pi}{2}-\arccos\frac{\cos\theta}{\sqrt{-\lambda}}\right)\left[\int_{\cos\theta}^{\infty} \hspace{-0cm}dr\,\frac{(r\cos\theta-r^2)^{2\Delta+1}}{r} +\int_{0}^{\infty} \hspace{-0cm}dr\,\frac{(-r\cos\theta-r^2)^{2\Delta+1}}{r} \right] \,.
\end{align}
We then have
\begin{align}\label{duesima}
    I=   &- \frac{\delta(0)\,(4|k_1|)^{2(\Delta+1)}}{4} \int_{-\frac{\pi}{2}}^{\frac{\pi}{2}}\hspace{-0.2cm}d\theta\left[ \left(\int_{-\frac{\pi}{2}}^{-\arccos\frac{\cos\theta}{\sqrt{-\lambda}}}\hspace{-1.0cm}d\theta'+ \int_{\arccos\frac{\cos\theta}{\sqrt{-\lambda}}}^{\frac{\pi}{2}}\hspace{-1.0cm}d\theta'\hspace{0.4cm}\right) \left(\int_{0}^{f_-} \hspace{-0.2cm}dr 
     +\int_{f_+}^{\cos\theta} \hspace{-0.4cm}dr\right) \frac{(r^2-r\cos\theta)^{2\Delta+1}\cos\theta'}{r\sqrt{\cos^2\theta'+4\frac{r}{\lambda} (\cos\theta-r)}} \right. \notag \\
    &\left.+2 \left(\frac{\pi}{2}-\arccos\frac{\cos\theta}{\sqrt{-\lambda}}\right)\left(\int_{\cos\theta}^{\infty} \hspace{-0cm}dr\,\frac{(r\cos\theta-r^2)^{2\Delta+1}}{r} +\int_{0}^{\infty} \hspace{-0cm}dr\,\frac{(-r\cos\theta-r^2)^{2\Delta+1}}{r} \right)\right] \,,
\end{align}

The double integrals of (\ref{duesima}) are solvable, but divergent. They can be suitably regulated as:
\begin{align}\label{di1}
   & -2\int_{-\frac{\pi}{2}}^{\frac{\pi}{2}}\hspace{-0.2cm}d\theta\left(\frac{\pi}{2}-\arccos\frac{\cos\theta}{\sqrt{-\lambda}}\right)\left(\int_{\cos\theta}^{\infty} dr\,(\cos\theta-r) -\int_{0}^{\infty} dr\,(\cos\theta+r) \right) \notag \\
   & =\lim_{L\to\infty}\left[ \pi^2 L^2+\frac{1}{4}\pi^2-2\int_{-\frac{\pi}{2}}^{\frac{\pi}{2}}\hspace{-0.2cm}d\theta\,\arccos\frac{\cos\theta}{\sqrt{-\lambda}}\left(\frac{1}{2}\cos^2(\theta)+L^2\right) \right]\,.
\end{align}

Then there are more complicated triple integrals; Integrating these in $r$ and adding (\ref{di1}) we obtain
\begin{align}\label{lambdamina}
F(\lambda,0)=   \sqrt{-\lambda}\int_{0}^{\frac{\pi}{2}}\hspace{-0.2cm}d\theta\int_{\arccos\frac{\cos\theta}{\sqrt{-\lambda}}}^{\frac{\pi}{2}}\hspace{-1.1cm}d\theta'\; \cos\theta'\cos\theta\,&\log\left(\frac{\frac{\cos\theta}{\sqrt{-\lambda}}+ \cos\theta' }{\frac{\cos\theta}{\sqrt{-\lambda}}- \cos\theta'}\right) \notag \\
&+\lim_{L\to\infty}\left[  \pi^2(L^2+\frac{1}{4})-4\int_0^{\frac{\pi}{2}}\hspace{-0.2cm}d\theta\,\arccos\frac{\cos\theta}{\sqrt{-\lambda}}\left(\frac{1}{2}\cos^2(\theta)+L^2\right) \right]\;.
\end{align}

Denoting $a=\frac{\cos\theta}{\sqrt{-\lambda}}$ the primitive in $\theta'$ of (\ref{lambdamina}) can be written as:
\begin{align}\label{prim1}
   f(\theta')\equiv & \int d\theta' \cos\theta'\,\log\left(\frac{a+ \cos\theta' }{a-\cos\theta'} \right) \notag \\
    &=2\sqrt{1-a^2}\left[\tanh^{-1}\left(\frac{\sqrt{1-a}}{\sqrt{1+a}}\tan(\theta'/2)\right)-\tanh^{-1}\left(\frac{\sqrt{1+a}}{\sqrt{1-a}}\tan(\theta'/2)\right)\right] +2\,a\,\theta'+\sin\theta'\log\left(\frac{a+ \cos\theta' }{a-\cos\theta'} \right)\,
\end{align}
where $f(\theta')$ at the points $\theta'=\pi/2,\arccos\,a$ is respectively:
\[
f(\pi/2)=2\sqrt{1-a^2}\left[\tanh^{-1}\left(\frac{\sqrt{1-a}}{\sqrt{1+a}}\right)-\tanh^{-1}\left(\frac{\sqrt{1+a}}{\sqrt{1-a}}\right)\right]+a\,\pi \,,
\]
and, by introducing a temporary regularization inside the divergent terms $\arccos a\rightarrow \epsilon+\arccos a$ with $1\gg\epsilon>0$,
\begin{align*}
f(\arccos a)&=2\sqrt{1-a^2}\left[\tanh^{-1}\left(\frac{1-a}{1+a}\right)-\tanh^{-1}\left(1+\frac{\epsilon}{\sqrt{1-a^2}}+O(\epsilon^2)\right)\right]+2a\,\arccos a \notag \\
&+\sqrt{1-a^2}\log\left(\frac{2 a }{\epsilon\sqrt{1-a^2}} +\frac{a^2}{a^2-1}+O(\epsilon)\right)
=\sqrt{1-a^2}\left[-\log\left(1-a^2\right) +i\pi \right]+2a\,\arccos a +O(\epsilon\rightarrow 0)\,.
\end{align*}
Then
 \begin{align}\label{lambdamin2ci}
&F(\lambda,0)= \lim_{L\to\infty}L^2\left(\pi^2-4\int_0^{\frac{\pi}{2}}\hspace{-0.2cm}d\theta\,\arccos a\right)+\frac{\pi^2}{4} -\lambda \int_0^{\frac{\pi}{2}}\hspace{-0.2cm}d\theta \;\left[a \left(f(\pi/2)-f(\arccos a)\right)  -2a^2\,\arccos a \right]\notag \\
 &=-\lambda \int_0^{\frac{\pi}{2}}\hspace{-0.2cm}d\theta \; \left[a\sqrt{1-a^2}\left(2\tanh^{-1}\left(\frac{\sqrt{1-a}}{\sqrt{1+a}}\right)-2\tanh^{-1}\left(\frac{\sqrt{1+a}}{\sqrt{1-a}}\right)+\log\left(1-a^2\right)-i\pi\right)+a^2\,\pi -4a^2\,\arccos a\right]\notag \\
&+\frac{\pi^2}{4}+\lim_{L\to\infty}L^2\left(\pi^2-4\int_0^{\frac{\pi}{2}}\hspace{-0.2cm}d\theta\,\arccos a\right)\notag \\
&=-\lambda \int_0^{\frac{\pi}{2}}\hspace{-0.2cm}d\theta \; a\left(\sqrt{1-a^2}\log\left(1-a^2\right)+a\,\pi-4a\arccos a \right)
+\frac{\pi^2}{4}+\lim_{L\to\infty}L^2\left(\pi^2-4\int_0^{\frac{\pi}{2}}\hspace{-0.2cm}d\theta\,\arccos a\right)\notag \\
&= (-\lambda)^{\frac{3}{2}}\hspace{-0.2cm} \int_{0}^{1/\sqrt{-\lambda}}\hspace{-0.8cm}da  \; \frac{a}{\sqrt{1+\lambda a^2}}\left(\sqrt{1-a^2}\log\left(1-a^2\right)+a\,\pi-4a\arccos a\right)
+\frac{\pi^2}{4}+\lim_{L\to\infty}L^2\left(\pi^2-4\sqrt{-\lambda}\int_0^{1/\sqrt{-\lambda}}\hspace{-0.2cm}da\,\frac{\arccos a}{\sqrt{1+\lambda a^2}}\right)\,.
\end{align}

These integrals are quite complicated to study, however some limits are relatively simple to perform. 

The simplest one is $\lambda=-1$ for the two point function at momenta $k_1,-k_1$. From the integral we have
\begin{align}
F(\lambda=-1)=\frac{1}{2}+\frac{\pi^2}{4}+\lim_{L\to\infty} \frac{\pi^2}{2}L^2
\end{align}
The corresponding function (\ref{ententropy}) then is :
\begin{equation}
    S(k_1,-k_1)=-c\frac{4}{3} \,\pi(1+\frac{\pi^2}{2}+ \lim_{L\to\infty}L^2) \,|k_1|^2\lim_{\epsilon^{R/I}\to 0}\delta(k_1^R\epsilon^I-k_1^I\epsilon^R) \;.
\end{equation}

Considering instead $-\lambda \rightarrow \infty$ we can expand in $a\rightarrow 0$ :
 \begin{align}
F(\lambda,0)&= (-\lambda)^{\frac{3}{2}}\hspace{-0.2cm} \int_{0}^{1/\sqrt{-\lambda}}\hspace{-0.8cm}da  \; \frac{a}{\sqrt{1+\lambda a^2}}\left(-\pi\,a+3a^2+O(a^4)\right)+\frac{\pi^2}{4}+\lim_{L\to\infty}L^2\left(\pi^2-4\sqrt{-\lambda}\int_0^{1/\sqrt{-\lambda}}\hspace{-0.2cm}da\,\frac{(\frac{\pi}{2}-a+O(a^3))}{\sqrt{1+\lambda a^2}}\right) \notag \\
&= \frac{2}{\sqrt{-\lambda}}\left(1+\lim_{L\to\infty}2L^2\right)+O(-\frac{1}{\lambda}) \,.
\end{align} 

The final result then is:
\begin{equation}
    S(k_1,\lambda(\rightarrow -\infty)k_1)= -\frac{1}{\sqrt{-\lambda}}c\frac{16}{3} \;\pi \left(1+\lim_{L\to\infty}2L^2\right)\,|k_1|^2\lim_{\epsilon^{R/I}\to 0}\delta(k_1^R\epsilon^I-k_1^I\epsilon^R) +O(-\frac{1}{\lambda})\;.
\end{equation}

Finally we can consider the case $-\lambda\rightarrow 0$. Here we cannot use directly equation (\ref{lambdamin2ci}) as this is valid in the region $\lambda\neq 0$. We instead can either start again from the very beginning and solve the delta function in $r$ using (\ref{solr}) and repeat all the calculations or, equivalently, simply apply the transformation rule (\ref{cond}) to obtain
\[
F_-(\lambda\rightarrow 0_-)=|\lambda\rightarrow 0_-|F(\frac{1}{\lambda}\rightarrow -\infty)=2(-\lambda)^{\frac{3}{2}}\left(1+\lim_{L\to\infty}2L^2\right)+O((-\lambda)^2)
\]
which implies
\begin{equation}\label{zerominus}
    S(k_1,\lambda(\rightarrow 0_-)k_1)= -(-\lambda)^{\frac{3}{2}}c\frac{16}{3} \;\pi \left(1+\lim_{L\to\infty}2L^2\right)\,|k_1|^2 \lim_{\epsilon^{R/I}\to 0}\delta(k_1^R\epsilon^I-k_1^I\epsilon^R) +O((-\lambda)^2)\;.
\end{equation}
\subsubsection*{Discontinuity at $\lambda=0$ and regularization}
A surprising feature is that there is a discontinuity at $\lambda=0$ between the entanglement entropies $S(k_1,\lambda(\rightarrow 0_+)k_1)$ and $S(k_1,\lambda(\rightarrow 0_-)k_1)$, that is when $k_2$ is at the origin. On one side we have $S(k_1,\lambda(\rightarrow 0_+)k_1)=-c\frac{2}{3}\;\pi^3 \;\delta(0)\, |k_1|^2$, on the other (\ref{zerominus}). This discontinuity is in fact due to  regularization issues, where different divergencies and zeros should compensate each other to be consistent in the two limits. To better understand the point let us compute the value at $\lambda=0$ directly. There are a few, in principle equivalent choices, to compute the result but producing apparently different results for different choices in regularizing integrals. For instance considering (\ref{integnew}) at $\lambda=0$ and $\Delta=0$ with $p_2\rightarrow p_1$ (that produces the cleanest result, aka with the lesser number of divergences) we get
\begin{equation}
I = \delta(0)\,4|k_1|^2\int drd\theta dr'd\theta'\,\frac{r(\cos\theta-r)^2}{r'|\cos\theta|} \sum_{i=\pm} \delta(r-r_{*,i})\,.
\end{equation}
The delta function solutions now are $r_*=0$ and $r_*=\cos\theta$, the latter for positive $\cos\theta$ only. So regularizing things:
\begin{align}
I = &\delta(0)4|k_1|^2\int_{\epsilon'}^{L'} dr'\frac{1}{r'} \cdot 2\pi \left[\int_{-\frac{\pi}{2}}^{\frac{\pi}{2}} d\theta \left(\frac{r(\cos\theta-r)^2}{|\cos\theta|} |_{r=\epsilon}+\frac{r(\cos\theta-r)^2}{|\cos\theta|} |_{r=\cos\theta-\epsilon}\right)+\int_{\frac{\pi}{2}}^{\frac{3\pi}{2}} d\theta \frac{r(\cos\theta-r)^2}{|\cos\theta|} |_{r=\epsilon} \right] \notag \\
&=\delta(0)\,4|k_1|^2(\log L'-\log \epsilon')  \cdot 2\pi \left[4\epsilon+\pi\epsilon^2 \right]\,.
\end{align}
We see here that $\epsilon$ should depends on $\lambda$ at first order, and so also $L'$ and $\epsilon'$, to be able to recover at $\lambda=0$, either as $\lambda\rightarrow 0_+$ or $\lambda\rightarrow 0_-$,  the two previous results. Similarly if we had chosen to start with $p_2$ inside the integral a factor of $|\lambda|^2$ would have appeared in front as well as additional divergences. Again a consistent regularization scheme should account for a continuous result. We do not work out in detail such a procedure an leave it for future work.

\subsubsection*{Physical origin of the divergences}
We have seen that divergences in the twist operator two point function occur in two different ways: on one side we have a common $\delta(0)$ for both the $\lambda>0$ and $\lambda<0$ cases, on the other we have a UV divergence at large momenta which happens only for $\lambda<0$. This second divergence appears at $r\rightarrow\infty$ (and $r_{*,\pm}\rightarrow\infty$), which corresponds to $k,k'\rightarrow\infty$; tracing back to equation (\ref{eq:main}) this implies $x_1,x_2\rightarrow 0$, that is when the positions of the \emph{real} space twist operators come to coincide at the origin. This happens only for negative $\lambda$ because the square root in the $r_{*,\pm}$ solutions of the $\delta(p_1+p_2)$ permits them to go to infinity for large $r$ only for negative $\lambda$.

\bibliographystyle{JHEP}
\bibliography{paper}

\end{document}